\documentclass[preprint,12pt]{elsarticle}

\usepackage{listings}
\usepackage{todonotes} 

\usepackage{mathtools}
\usepackage{blkarray, bigstrut}
\usepackage{adjustbox}
\usepackage{verbatimbox}
\usepackage{xspace}

\usepackage{amssymb}

\usepackage{amsthm}

\usepackage{graphicx}
\usepackage{subfigure}

\usepackage{hyperref}

\usepackage{amsmath}
\usepackage{xcolor}
\usepackage{blkarray}

\usepackage{algpseudocode}
\usepackage[linesnumbered,ruled,vlined]{algorithm2e}

\usepackage{verbatim}
\usepackage{fancyvrb}

\usepackage{adjustbox}

\usepackage{verbatimbox}

\usepackage{nicematrix}

\DontPrintSemicolon

\definecolor{delim}{RGB}{20,105,176}
\definecolor{numb}{RGB}{106, 109, 32}
\definecolor{string}{rgb}{0.64,0.08,0.08}

\lstdefinelanguage{json}{
    numbers=left,
    numberstyle=\small,
    frame=single,
    rulecolor=\color{black},
    showspaces=false,
    showtabs=false,
    breaklines=true,
    postbreak=\raisebox{0ex}[0ex][0ex]{\ensuremath{\color{gray}\hookrightarrow\space}},
    breakatwhitespace=true,
    basicstyle=\ttfamily\small,
    upquote=true,
    morestring=[b]",
    stringstyle=\color{string},
    literate=
     *{0}{{{\color{numb}0}}}{1}
      {1}{{{\color{numb}1}}}{1}
      {2}{{{\color{numb}2}}}{1}
      {3}{{{\color{numb}3}}}{1}
      {4}{{{\color{numb}4}}}{1}
      {5}{{{\color{numb}5}}}{1}
      {6}{{{\color{numb}6}}}{1}
      {7}{{{\color{numb}7}}}{1}
      {8}{{{\color{numb}8}}}{1}
      {9}{{{\color{numb}9}}}{1}
      {\{}{{{\color{delim}{\{}}}}{1}
      {\}}{{{\color{delim}{\}}}}}{1}
      {[}{{{\color{delim}{[}}}}{1}
      {]}{{{\color{delim}{]}}}}{1},
}

\lstdefinelanguage{json2}{
    frame=single,
    rulecolor=\color{black},
    showspaces=false,
    showtabs=false,
    breaklines=true,
    postbreak=\raisebox{0ex}[0ex][0ex]{\ensuremath{\color{gray}\hookrightarrow\space}},
    breakatwhitespace=true,
    basicstyle=\ttfamily\small,
    upquote=true,
    morestring=[b]",
    stringstyle=\color{string},
    literate=
     *{0}{{{\color{numb}0}}}{1}
      {1}{{{\color{numb}1}}}{1}
      {2}{{{\color{numb}2}}}{1}
      {3}{{{\color{numb}3}}}{1}
      {4}{{{\color{numb}4}}}{1}
      {5}{{{\color{numb}5}}}{1}
      {6}{{{\color{numb}6}}}{1}
      {7}{{{\color{numb}7}}}{1}
      {8}{{{\color{numb}8}}}{1}
      {9}{{{\color{numb}9}}}{1}
      {\{}{{{\color{delim}{\{}}}}{1}
      {\}}{{{\color{delim}{\}}}}}{1}
      {[}{{{\color{delim}{[}}}}{1}
      {]}{{{\color{delim}{]}}}}{1},
}

\lstdefinelanguage{ini}{
    basicstyle=\ttfamily\small,
    columns=fullflexible,
    morecomment=[s][\color{purple}\bfseries]{[}{]},
    morecomment=[l]{\#},
    morecomment=[l]{;},
    commentstyle=\color{gray}\ttfamily,
    morekeywords={},
    otherkeywords={=,:},
    keywordstyle={\color{red}\bfseries}
}

\newcommand{\uiblong}{University of the Balearic Islands}
\newcommand{\uibacro}{UIB}
\newcommand{\github}{https://github.com/ocsix6/mobility-analysis-corporate-wifi}

\newcommand{\cslzero}{$\left< lv0, -, - \right>$}
\newcommand{\cslone}{$\left< lv1, bldgAT, - \right>$}
\newcommand{\csltwo}{$\left< lv2, bldgAT, 0flE \right>$}

\SetKwComment{Comment}{\color{green!50!black}// }{}
\SetKwProg{Function}{function}{}{}

\SetAlFnt{\scriptsize}

\journal{Journal of King Saud University - Computer and Information Sciences}

\begin{document}

\begin{frontmatter}



\title{Analysis of wireless network access logs for a hierarchical characterization of user mobility}


\author{Francisco Talavera\corref{}}
\ead{f.talavera@uib.es@uib.es}

\author{Isaac Lera\corref{}}
\ead{isaac.lera@uib.es}

\author{Carlos Guerrero\corref{mycorrespondingauthor}}
\ead{carlos.guerrero@uib.es}
\cortext[mycorrespondingauthor]{Corresponding author}

\address{Crta. Valldemossa km 7.5, Palma, E07121, SPAIN}

\address[mymainaddress]{Computer Science Department, University of Balearic Islands}

\begin{abstract}

 This paper presents a method that generates a hierarchical user mobility model from the analysis of the data available from Wi-Fi connections. The data obtained from the Wi-Fi infrastructure is defined in terms of the coverage areas of the access points that the users move through. These access points are recursively grouped into different levels of granularity based on their geospatial features. The track of a user is defined as a sequence of Wi-Fi access points, which is enough to simulate user mobility in, for example, fog scenarios. The hierarchical definition of the region under study is proposed to reduce the complexity of the model in high-scale scenarios and to increase the adaptability between scenarios with different geospatial features. The model creation is based on a user profiling method that uses a clustering algorithm and each user type is defined with a transition matrix between coverage areas and a time length vector for the areas. The method is applied to the case of the campus of the \uiblong. From the analysis of the mean square error of the results, we determined that the proposed method obtains good results for the transition matrices, but that the time vector definition should be improved.  The results also show lower complexity in the case of the hierarchical model, with one area for each building and three levels, in regard to a non-hierarchical model, with only one area and one level for the whole campus. 

\end{abstract}


\end{frontmatter}



\section{Introduction}

The number of research studies in the field of fog computing has increased significantly during the last years. In most of the cases, the experimental phase  and the testing of the new research proposals  are performed in simulation environments, mainly because of the difficulties to access real infrastructures with real users. Simulated results are only reliable if the simulation uses a model based on a real scenario. Thus, simulations typically include synthetic traces created from a model defined from real data~\cite{842273}.  But a quick review of the up-to-date literature in the field of fog computing~\cite{10.1002/spe.2766, OGUNDOYIN2021100937} shows that most of the experiments are performed with random non-realistic models, because of the difficulties to access real data from these environments to create the models.


 Moreover, fog environments are strongly influenced by the mobility of the users in the system, contrary to other traditional distributed architectures. Consequently, user mobility models, based on real-data, are also necessary to obtain reliable results in the simulated experiments.   
User mobility influences fog architecture concerning the access points (APs) to which the users are connected to. In other words, the connection of a user to a given AP determines, for example, the origin of the requests, the number of requests sent from each AP, the network route the requests and responses are transmitted along, the network load, etc.  Thus, in a simulation, the fog users' mobility model determines the APs where the users are connected to. This model is not influenced by the underlying network technology, and these connections could be established with, for example, a Wi-Fi AP, a 5G radio access network, etc. without altering the mobility patterns of the users.

The importance of user information for technological companies compromises its publication in open data catalogues, and user mobility is one such example. There are many studies in the literature that require of user mobility and that illustrate the importance of having user mobility models~\cite{surveyToch2019}. There are examples of the key aspect of this user mobility modeling in technological research areas --such as mobile networks~\cite{622908,8570749}, location-based applications~\cite{6413812}, intelligent transport systems~\cite{doi:10.1177/1550147720963751}, cloud performance~\cite{7514946}, or fog infrastructures~\cite{7912261}-- and in social areas --such as crowds management~\cite{HUANG2018147}, travel patterns discovering~\cite{Hoogendoorn2005}, natural areas management~\cite{MEIJLES201444}, or analysis of sport activities~\cite{10.1371/journal.pone.0177712}--.

Modeling methodologies to characterize user mobility still have room for new proposals.  One example is the case of our domain problem, in which the mobility of the user is  modeled hierarchically and  created from a small set of mobility data, only requiring the data related to the APs in which the users are connected to.

In this work, we present a method to create a  hierarchical  user mobility model that recursively defines geospatial levels by grouping the neighboring APs into regions and zones of different granularity. The input data of the method is a data set of user connections to the APs of a Wi-Fi infrastructure. We focus our study on the use of Wi-Fi technology because this type of data is more accessible than 5G data, which requires the collaboration of public communication companies. But our method could be easily applied in other types of connection networks. 

 Note that we do not propose a new system or tool for user location or for gathering this data. We use such a type of systems to gather user location data, which are subsequently used by our proposed method for the creation of the user mobility models. In our particular case study, we decided to use Aruba Location Engine (ALE) because it was already deployed in our infrastructure, but any other system, such as Cisco Prime Infrastructure would be also suitable to gather the data.


The paper is organized as follows:  in the remainder of this introductory section, the motivations and contributions of the proposed work are highlighted; Section~\ref{sect_relatedwork} reviews related research works;   Section~\ref{sect_probstatement} introduces the context of user mobility modeling; Section~\ref{sect_proposedsolution} includes the details of the proposed method for user mobility modeling; Section~\ref{sect:casestudy} presents the application of the proposed method in the study of the user mobility in the campus of the \uibacro, and the results are analyzed; and finally, Section~\ref{sect_conclusions} summarizes the conclusions and establishes future research lines. 




\subsection{Motivation and contributions}

The evaluation of new fog proposals is usually performed in a simulator in their early experimental phases, because of the large scale of these infrastructures. Simulations need real data to obtain reliable results. But the number of open data sets related to user mobility in IT infrastructures is very limited~\cite{surveyLuca2021}, and most of them do not include the relationship with the computing infrastructures~\cite{electronics9040560}. 

Obtaining data from real scenarios is complex and many researchers do not have access to suitable data sets, nor to infrastructures where they can collect the data.

From the analysis of the experiments of most of the research works~\cite{10.1002/spe.2766, OGUNDOYIN2021100937}, it is observed that there is an important challenge related to the use of real models in simulations. When researchers of fog infrastructures need to test and validate their proposals, they need to face two important challenges, depending on the resources they have. 

The first case is when the researchers have access to the deployed infrastructures they want to simulate, and they can collect users' mobility data and their interactions with the computing infrastructure. In those cases, the main challenge is that the computational complexity to create the user mobility model is very high.

The second case is when the researchers can only access open data sets or models, whose availability is very limited, and the existing models need to be adapted to the infrastructure and the geospatial features of their study case.

Consequently, two research questions arise from these research challenges:

\begin{itemize}
    \item RQ1: Is it possible to model user mobility in such a way that allows us to extrapolate/adapt a model between case studies with different geospatial features?
    \item RQ2: Is it possible to reduce the complexity of the mobility creation for large scale scenarios, such as fog infrastructures?
\end{itemize}

We have analysed several up-to-date reviews and surveys in the field of user mobility modeling, as we comment in Section~\ref{sect_relatedwork},  and to the best of our knowledge, current user modeling proposals lack both challenges, the computational complexity of the mobility model creation for large scale domains, and the adaptation of the mobility models between scenarios with different geospatial features.

Our proposal addresses both challenges by combining approaches from the fields of user characterization and geospatial studies. We combine the use of transition matrices to model user mobility behavior~\cite{BarbosaFilho2018HumanMM} and the definition of a hierarchical geospatial organization~\cite{10.1145/2674918.2674921,articleXu2015} of the networking resources in which the users are connected to.  We concrete our proposal in the following research hypothesis: User mobility can be modeled by defining hierarchical levels of the geospatial zones of the region under study, and modeling  mobility at  each of these levels with transition matrices between the zones. This hierarchical approach reduces  complexity and increases the extrapolation of the resulting model.

The great volume of data regarding user connections requires a data analysis solution that reduces the problem size without compromising results. With the use of a geospatial hierarchical decomposition of the region under study into levels and zones, the modeling process can be split into smaller sets, reducing the computational complexity.

Additionally, the hierarchical definition of the mobility model allows us to easily adapt it to scenarios with different geospatial features by scaling it up/down. For example, zones with given features can be duplicated or removed to adapt the mobility model to other case studies.


We particularize our proposal in the case of Wi-Fi infrastructures and apply it in a real study case. Thus, the contributions of this paper are:

\begin{itemize}
    \item A method for hierarchical user mobility modeling. The model is defined with pairs of a stochastic transition matrix and a time vector, that respectively model the changes in the coverage areas of the APs (i.e., the AP that the user is connected to) and the time that the user stays in that coverage area. The input of the model is the connection data obtained from Wi-Fi APs. The proposed method can be easily extended to other connection technologies, for example 5G networks. 
    \item The application of the proposed method to a real scenario (the campus of the \uiblong, \uibacro, a medium size university) to validate and test the proposed method. The obtained model has been also published in an open repository to allow  other researchers to use/adapt/extend the model to their specific necessities.
\end{itemize}

We have highlighted the use of mobility data in the evaluation of fog infrastructures and we have focused on the applicability of our method to mobility generation in this type of environment. But our mobility models can be used in any type of evaluation/experiment that requires simulating the user mobility in a wireless environment. The only constraint is that the user mobility is modeled and expressed in terms of the APs where the users are connected to, i.e., the coverage area of the network access devices.

\section{Related work}
\label{sect_relatedwork}






{

The study of human mobility data plays a nuclear function in understanding scientific areas related, as in our case, the evaluation of fog infrastructures. But it is not less important its role in fields such as culture, spread of epidemics, environmental impact, tourism industry. There is an extended diversity of studies on the analysis of human mobility in so many scientific areas, reflecting its importance. Many surveys collect and classify these researches, generally based on the data set type and the purpose of the study~\cite{surveyToch2019,surveyLuca2021,BarbosaFilho2018HumanMM,10.1145/3308560.3320099,10.1145/3347146.3359110,king2021survey,RWWang2019,Solmaz2019,Thornton2018HumanMA,Hess2016}.


Becker et al.~\cite{becker2013human} demonstrated the value of cellular
network data for human mobility modeling. But their analysis of the mobility was focused on geographic aggregation of the results, instead of a single user-aware modeling. Consequently, traces of user movement can not be obtained from their mobility characterization.

Azebedo et al.~\cite{4917569} analyzed mobility real traces obtained with GPS technology. They proposed to base the mobility characterization in the study of the probability and the cumulative distribution functions of the velocity, acceleration, direction angle change and pause time of the user traces. They identified that  velocity  and  acceleration  components  follow a  Normal  distribution and that  the  direction  angle  change  component and the pause time are better represented by a Log-normal. This results are very valuable to generate synthetic traces of users in general scenarios. But for our target, emulating the user connections to APs, a more detailed model is required.

Gu et al.~\cite{7926085} dealt with the complexity of the model creation by proposing a distributed method for the mobility modeling. They propose a distributed version of the centralized algorithm NN-K-SVD. But the resulting model is not suitable for the emulation of user connections to APs.

Thuillier et al.~\cite{8014487} characterized user mobility by first profiling the users with a the k-mean clustering algorithm. Although our proposal also creates user clusters, in our case the clustering group users with similar mobility. On the contrary, Thuillier et al. based the clustering in the features of call detail records.


In the last decades, data from GSM network records have allowed the characterization of mobility patterns~\cite{BarbosaFilho2018HumanMM} with a greater volume of samples than other technologies i.e. GPS. Following this ubiquitous technological evolution, phones' Wi-Fi adapters describe in-detail indoor spatio-temporal resolutions, which were difficult to obtain with previous technologies, and also provide a large volume of samples. Any Wi-Fi device sends beacon-like messages to Wi-Fi APs which are gathered in the AP for logging purposes.

Traunmueller et al.~\cite{TRAUNMUELLER20184} presented a study focused on improving urban management and planning decisions using Wi-Fi probes. They used 54 APs in Lower Manhattan over one week getting 30 million observations from 800.000 unique devices. After data anonymization and cleaning processing, they conducted a graph analysis modeling the AP locations as the nodes, and the user movement between consecutive APs as the edges. They generated a street usage intensity network model and paths of travel at a time.  They used the model to interpret pedestrian routes and frequencies of points of interest (ferries, buildings, street connectors, etc.). Our study differs in that it is mainly focused on the method to obtain the mobility model, instead of the obtained mobility model. In any case, our case study, in a comparative scale, uses 425 APs located on floors and rooms in 18 buildings, resulting in 633000 filtered trajectories recorded in one week. There are other more simplistic studies which do not consider the user paths, and they are also focused on the Wi-Fi data set instead of the modeling method.

Uras et al.~\cite{Uras2019} conducted another human mobility study to identify human density, flows, patterns and heat maps through Wi-Fi probe data.  The experiment setup was on three locations: a street in Turing,  the Alba center, and the University of Cagliari where they deployed a small number of  APs in comparison with other studies ---1, 5 and 8 devices respectively---. They simplified the statistical methods by setting seven time periods and counting the permanence time of unique devices in the AP radius. This study shows the flexibility of these logs to obtain mobility indicators and ad hoc models. But the method they used to create the model is very limited for large scale scenarios.

Another important issue with these data sets is the estimation of the number of mobile devices present at a certain place and time and the number of people. Oliveira et al.~\cite{Oliveira2019} focused on this problem using several threshold values with Pearson coefficient correlations. The article of Balzotti et al.~\cite{BALZOTTI201825} uses mobile phone data provided by an Italian telecommunication company to adjust the length of the movement flow. Hoteit et al.~\cite{hoteit2017} proposed a technique to reduce the error between real and estimated human trajectories and to identify the period of time where users’ locations remain undefined. We addressed this problem using an experiment threshold since our context offers less noise in this undefined user state.

Gao et al.~\cite{10.1145/1860093.1860103} based mobility modeling in the use of Hidden Markov Models obtained from data of Wi-Fi infrastructures. But their goal was to obtain a fine-grained model (a higher precise user location), instead of a course-grained model (determined by the AP location or the coverage area). Because we are interested in the generation of user emulation in fog environments, course-grained models are more suitable. 

Additionally to the work of Gao et al., Markov modeling is studied in an important number of researches for modeling user mobility, most of them using GPS traces. Examples of variants for the Markov modeling are: simple Markov model using GPS traces~\cite{1167224}; semi-Markov model using GPS, GSM and Wi-Fi datasets~\cite{6199868}; hidden Markov model using GPS traces~\cite{10.1145/2370216.2370421}; mixed Markov model using GPS traces~\cite{10.1145/2093973.2093979}; mobility Markov chain using GPS traces~\cite{10.1145/2181196.2181199},  extended mobility Markov chain using call detail records~\cite{6757228}, variable-order Markov model using GPS traces~\cite{yan2013}, hidden semi-Markov model~\cite{YU2003235}, or spherical hidden-Markov model using geotagged Twitter datasets~\cite{zhu2018spherical}.

Our proposal of modeling user mobility with transition matrices is supported by all these related references. Although they cover a wide range of alternatives, to the best of our knowledge, we are the first work that considers a hierarchical modeling based on transition matrices by fixing the user position in terms of the APs where they are connected to.

Finally, we cite some related studies that implemented similar spatio-temporal aggregations with the objective of focusing the study on a specific region without losing information on the context of the subsumed regions.

Xie et al.~\cite{10.1145/2674918.2674921} implemented a histogram tree to identify the region through which a vehicle circulates. Xu et al.~\cite{articleXu2015} apply a hierarchical decomposition of the phone location data set from Shenzhen, China, in mobile phone towers. Both studies obtain a better performance analysis using this hierarchical decomposition but in our case, the decomposition is used to describe the flow model not only to facilitate the operative. It is a vision of the movement with different granularity levels. 


}

\section{Problem statement}
\label{sect_probstatement}

In general terms, users' movements are represented as a temporal sequence of geographical positions. Often the trajectories are recorded by GPS devices that offer fine-grained accuracy~\cite{10.1371/journal.pone.0177712}. However, other technologies, and their infrastructures, make the capture of the passage of users over regions possible, by associating personal devices and bind points. Phones and cellular radio towers are an example or, as in our case, Wi-Fi devices and APs. 


It is observed that mobility modeling usually has a high computational complexity~\cite{fulop2009accurate}, and the generated models are particular for specific scenarios and they lack in adaptability to other scenarios with different geospatial features~\cite{xie2013survey}. 
We propose a geospatial hierarchy user mobility modeling to address the problem of the computational complexity of large scale mobility scenarios and the low capacity of adaptation of the obtained models. Our model defines the interactions between users and network access devices, and the evolution of these interactions, i.e., the model represents the sequence of APs the users are connected to during their sessions/tracks.

\begin{figure}[!t]
\centering
\includegraphics[width=0.98\textwidth]{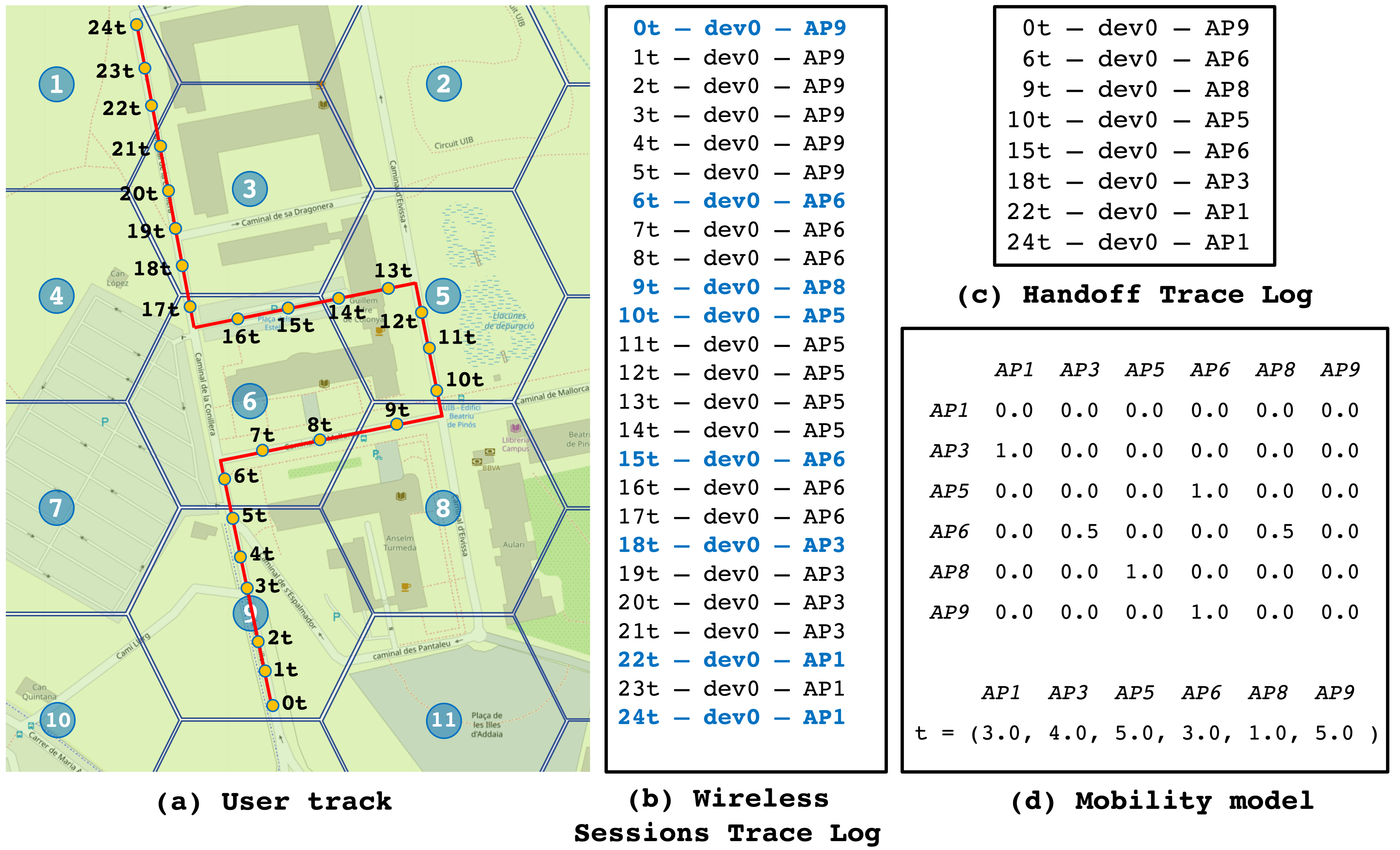}
\caption{Example of a user track and the corresponding mobility model.}
\label{fig_usertrack}
\end{figure}

Figure~\ref{fig_usertrack}.a shows an example of an user\footnote{ We use the terms user and device indistinctly along the paper. They refer to the concept of \textit{thing}, in the context of the Internet of Things, an entity that requests a service or generates and transfers data.} trajectory in the university campus of the \uibacro. The user follows the red line path over the background campus image, that also includes the coverage areas --represented with hexagons-- and the AP of each area --represented with a numbered blue dot in the center of the hexagon--. This trajectory also follows a time distribution --the orange dots-- that synthesizes the stay length in that space region. 

From the fog infrastructure perspective, the trajectory depends on the communication probes between the user devices and Wi-Fi APs. These probes depend on the coverage mesh and the messages that both entities exchange. In any case, the traffic link, between the device with the AP in a point in time, enables the characterization of the user movement. Thus, we can create an access log that includes 3-tuples lines relating to time, device/user and AP. We name this type of data set as wireless session access log. The session of a given user corresponds to the sub-set of samples with the same device identifier. Generally, the wireless session access log interleaves samples of different user sessions. Figure~\ref{fig_usertrack}.b shows the wireless session access log from the example of Figure~\ref{fig_usertrack}.


The key samples of a user session can be even reduced to the first and the last ones (that indicates the starting and ending points) and the first one of each new AP connection (blue colored lines in Figure~\ref{fig_usertrack}.b). The other samples are only related to the accuracy of the samples and they do not provide any additional information. With this subset of samples, the handoffs\footnote{A handoff happens when a device changes the AP where it is connected to.} of the user connections are identified. We name the handoff trace log to this data subset that only includes the filtered wireless sessions. Figure~\ref{fig_usertrack}.c shows the handoff trace log including a single filtered wireless connection session, of the example in  Figure~\ref{fig_usertrack}.

Our proposal represents the user mobility model, as in other previous studies, with a stochastic transition matrix (the Origin-Destination matrix~\cite{BarbosaFilho2018HumanMM}), that represents the probability of AP handoffs. We extend that model with the inclusion of a time vector that represents the average length time of the connection of a user to a given AP.  Figure~\ref{fig_usertrack}.d shows the transition matrix and the time length vector of the example.

We also extend the user modeling by defining a geospatial hierarchy of the network devices where the users are connected to. 
The hierarchical modeling is performed by grouping the neighboring APs in different levels of granularity, by taking into account the geospatial organization of the region under study. When this geospatial organization is taken into account, the adaptation of the resulting model to regions with different features will be easier by duplicating zones with similar features or by removing the ones without similarities.

\begin{figure}[!t]
\centering
\includegraphics[width=0.98\textwidth]{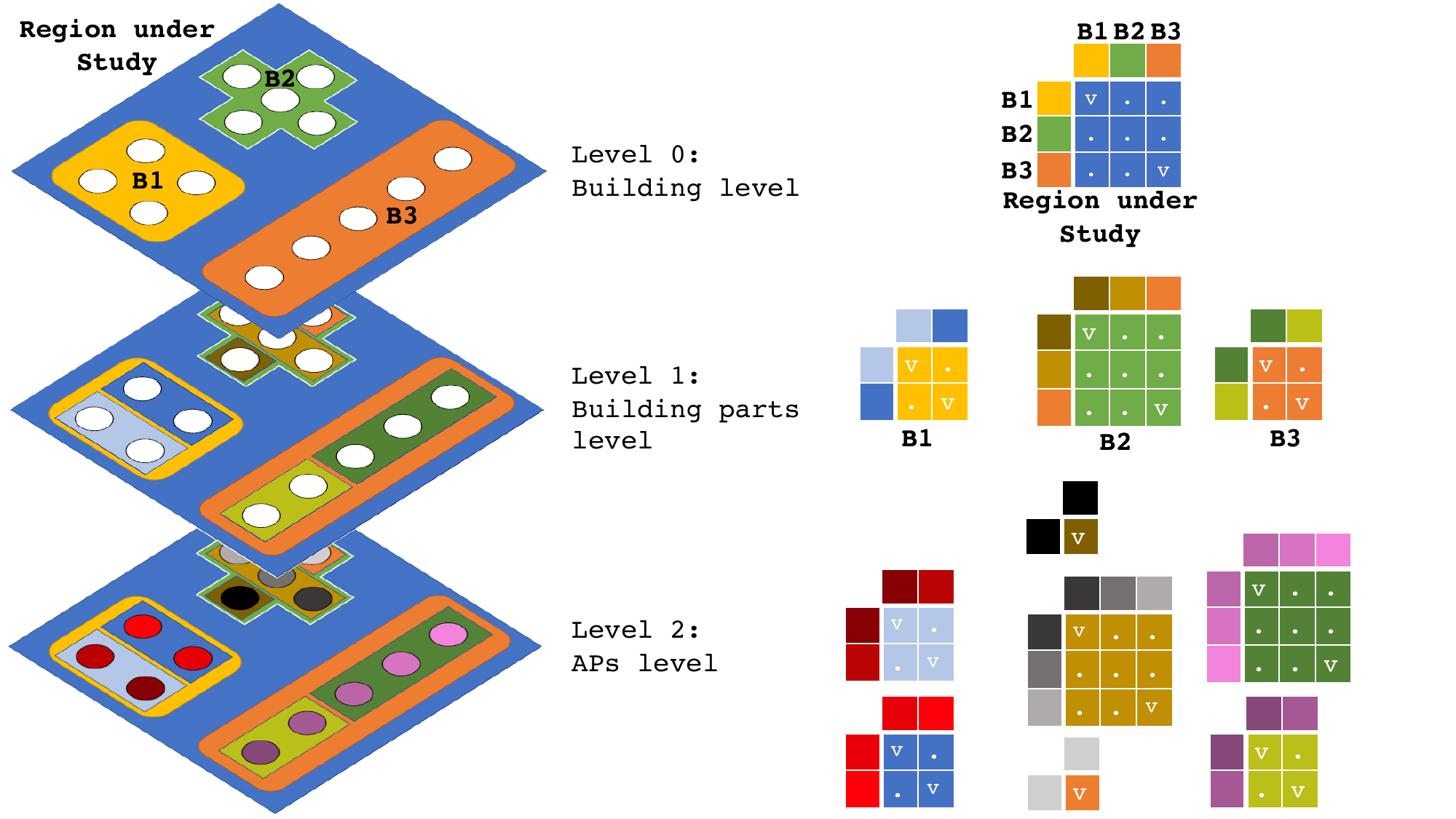}
\caption{Example of geospatial hierarchical modeling of user mobility.}
\label{fig_geospatialhierarchy}
\end{figure}

For example, in a first level, the groups could be created with the APs in the same building. In a second level, a set of APs in a building could be split into the parts of the building (for example, floors or wings), and this would be repeated recursively until the granularity reaches only one AP per group.
For example, Figure~\ref{fig_geospatialhierarchy} shows a specific case of three levels of granularity.

This recursive geospatial division of the region is directly reflected in the mobility model because the model is also split into independent transition matrices for each geospatial zone and each level. This is also shown in Figure~\ref{fig_geospatialhierarchy}, where the first level is represented with a matrix of the size of the number of buildings, that models the movement between buildings in the region under study. In the second level, a new matrix for each of the buildings is created to model the movement inside each of them. The matrix size corresponds to the number of parts a building is divided into.

Once the researchers have a mobility model, they need to adapt it to a specific problem definition to generate specific synthetic traces from a general mobility model. A simple and general adaptation refers to changing the percentages of each user type, the number of user types, or the user speed, between others. But if geospatial features need to be adapted, non-hierarchical models can not be adapted easily. In those cases, the use of hierarchical geospatial models simplified the adaptation process.

\begin{figure}[!t]
\centering
\includegraphics[width=0.98\textwidth]{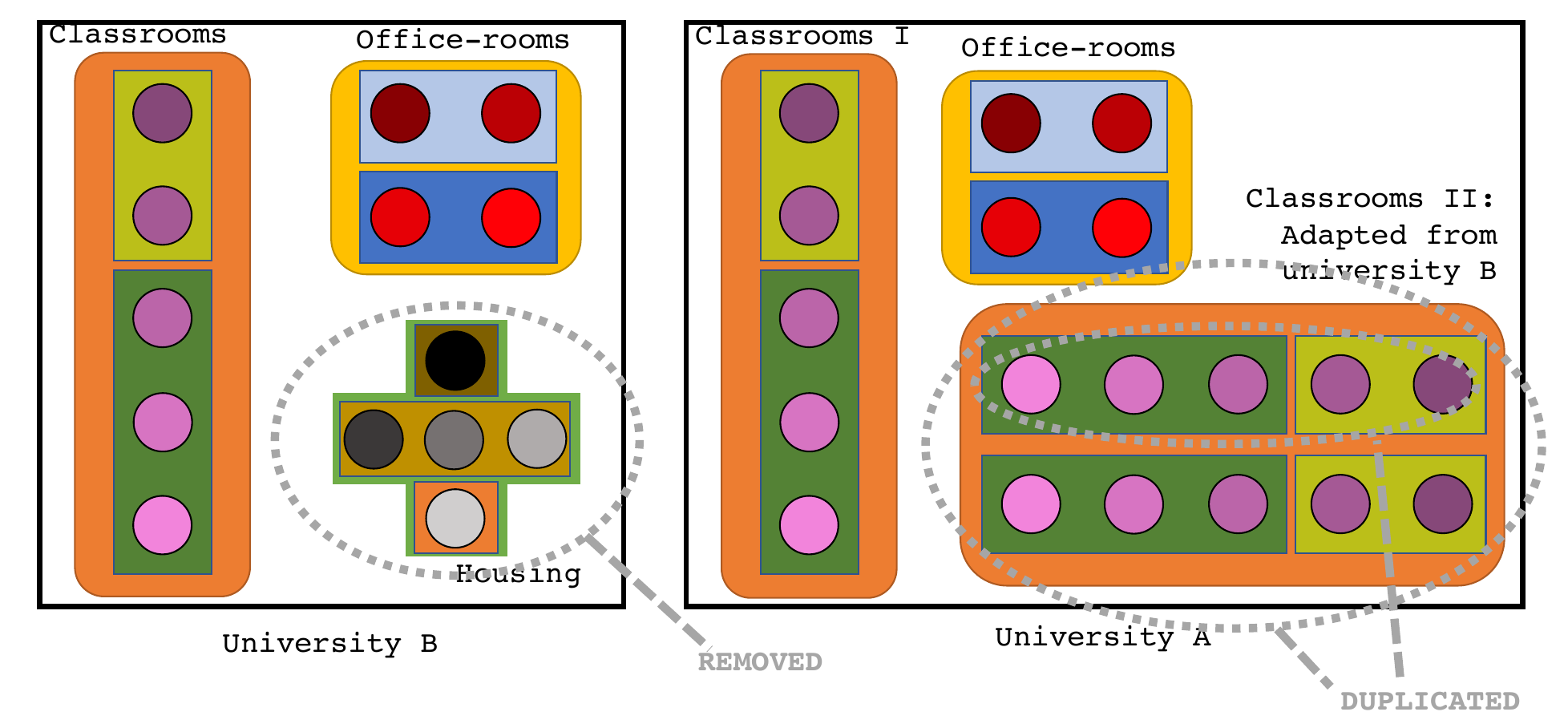}
\includegraphics[width=0.98\textwidth]{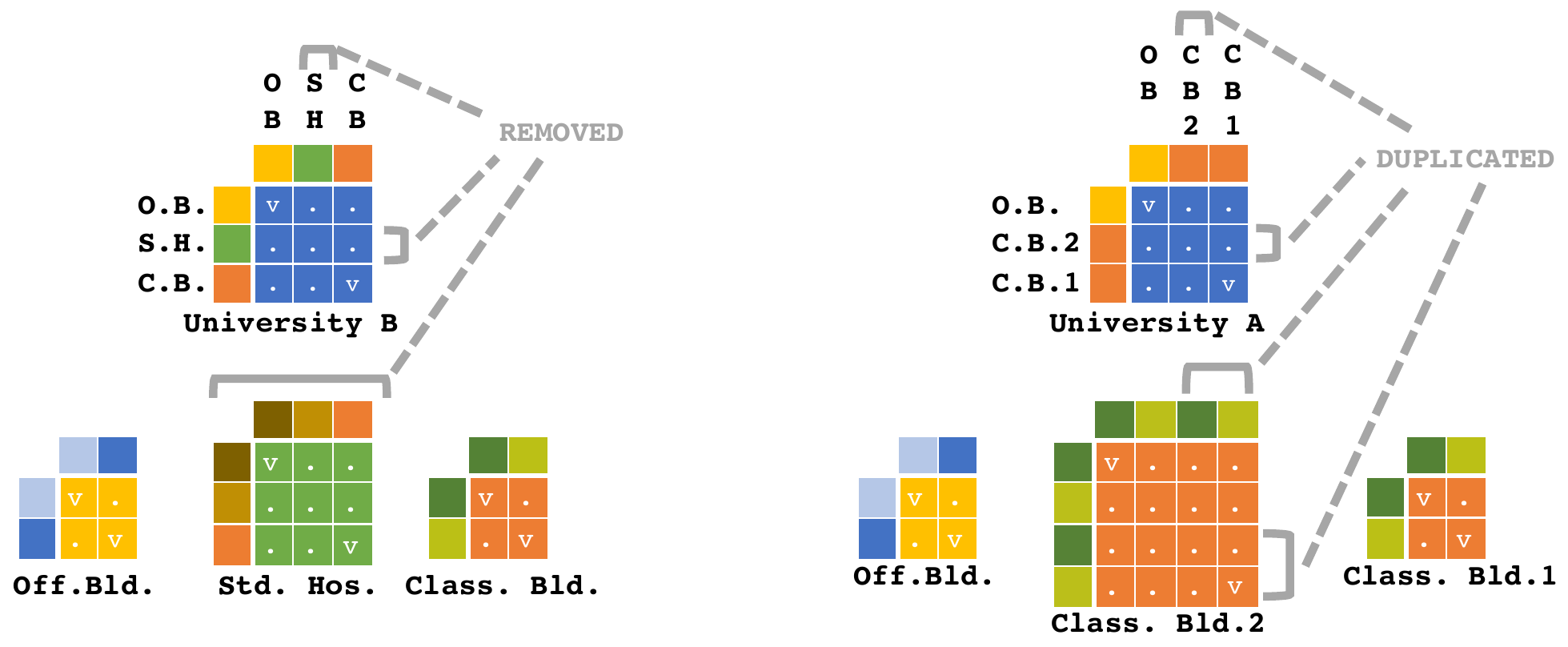}
\caption{Example of model adaptation between scenarios with different geospatial features.}
\label{fig_adapt_ex}
\end{figure}

Figure~\ref{fig_adapt_ex} shows an example in the context of the study case in Section~\ref{sect:casestudy}, a university campus. Imagine that a group of researchers wants to test the performance of a new fog infrastructure simulating the movement of the users in their campus (university A).  These researchers do not have access to real data from the students' movement and they decide to adapt a mobility model from another university that is available in an open repository (university B).

The university B model is defined with our hierarchical proposal and it models a campus with one classroom building (with two zones inside the building), one teachers' offices building (with two zones inside the building), and one student housing (with three zones inside the building). The problem is that university A has different geospatial features, with two classrooms buildings (one of them with similar features to University B and the other one with the double size), one teachers' building (with similar features to university B) and no student housing. Consequently, the mobility model of university B is not directly applicable to university A. 

Because of the hierarchical definition of the mobility model, the researchers can adapt the model of university B by removing the student housing, duplicating the classroom building in the building level of the hierarchy, and by increasing the size of this second duplicated classroom building by doubling the zones in the building-parts level. These adaptations are graphically represented in Figure~\ref{fig_adapt_ex}. Additionally, the figure also shows how the transition matrices are modified, by also duplicating or removing matrices and columns/rows of those matrices.



\section{Proposed solution}
\label{sect_proposedsolution}

\begin{figure}[!t]
\centering
\includegraphics[width=0.9\textwidth]{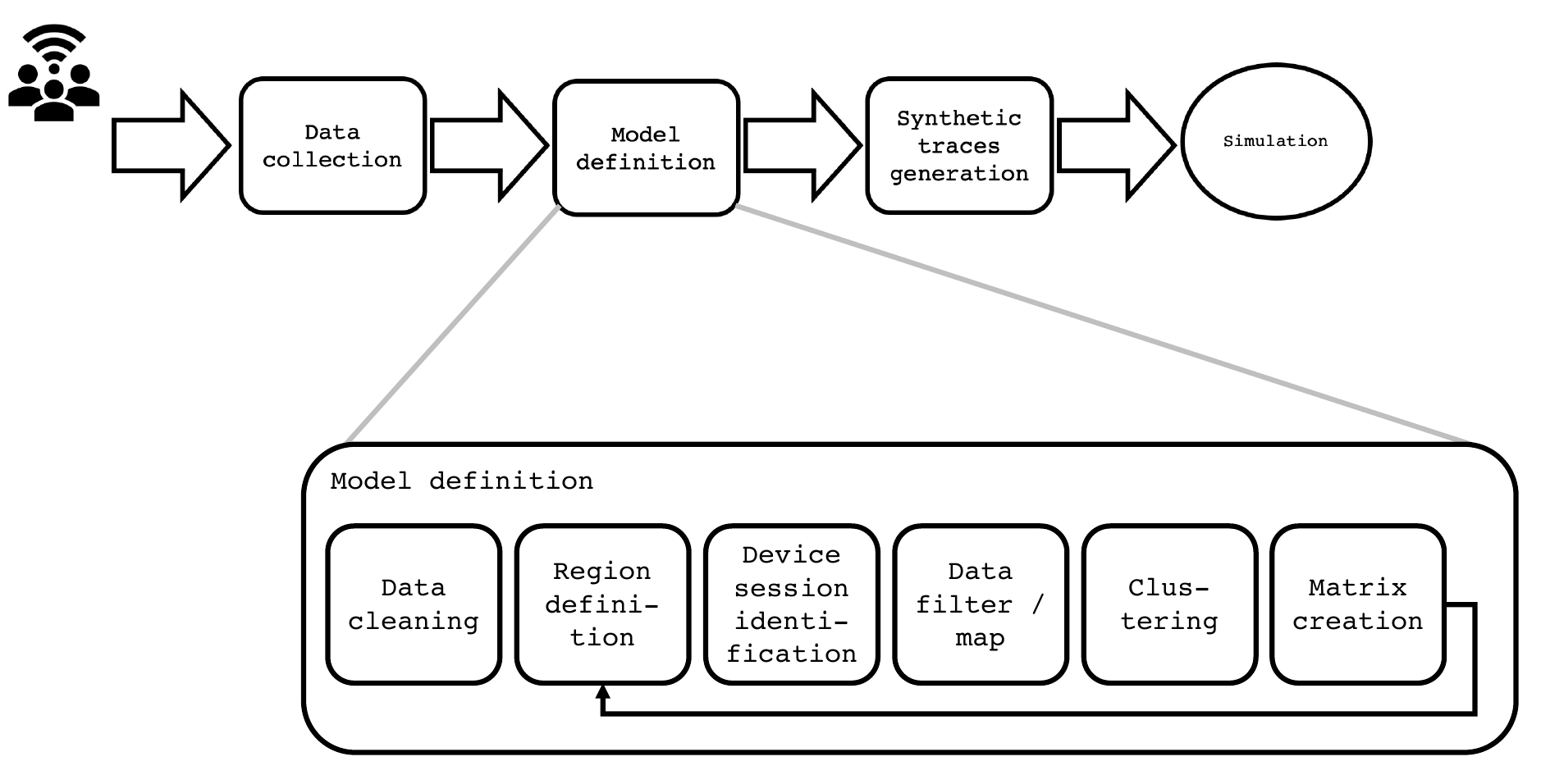}
\caption{Complete life-cycle of our proposed method for the hierarchical modeling of user mobility based on transition matrices.}
\label{fig_stepsmobilitymodel}
\end{figure}

This section presents the details of our proposed method that covers the requirements to address the research problem stated in the previous section (Section~\ref{sect_probstatement}). Our method involves successive phases of data collection, the definition of the model, and generation of synthetic data (Figure~\ref{fig_stepsmobilitymodel}).  By this, researchers are able to create synthetic traces with the mobility model obtained from these phases. Synthetic traces of user mobility can be exported to simulation/emulation engines for a more realistic evaluation of, for example, fog environments.

The phases in the method we propose are based on the traditional problem of characterizing e-business workloads with Customer Behavior Model Graphs (CBMG)~\cite{10.5555/556438}. But we have extended this method to incorporate a geospatial hierarchy of the APs that: (a) facilitates the adaptation of the models between scenarios with different geospatial features; (b) reduces the computational complexity of the generation of the model.

In summary, workload characterization with CBMG uses the web user requests stored in the web access logs to create users characterizations based on their web browsing. To do that, the user sessions are firstly identified, because the end of the user sessions is not registered in the weblogs. Subsequently, a clustering algorithm is executed to identify different types of users, and each user session is associated with one user type. Finally, one web transition matrix is generated for each user type using the set of user sessions of that user type.

We propose to revisit this method from e-business workload characterization~\cite{10.1145/1071021.1071039} to be applied to the problem domain of user mobility in fog architectures.   In our research problem, we deal with user connections to APs instead of web user requests.

One of the new challenges of applying or adapting this method to fog user mobility  is the size of the problem. The number of web pages mapped to states is usually much smaller than the number of APs in a region under study.  Our proposal solves this problem and the following subsections explain the adaptation of the life-cycle of this method (Figure~\ref{fig_stepsmobilitymodel}) for the user mobility modeling.


\subsection{First phase: data collection}

Our problem domain only needs to characterize AP connection length and the handoffs to represent the user mobility. Consequently, the first phase, data collection, is in charge of monitoring the user mobility by collecting the wireless sessions trace logs that characterize the user tracks. 

The most straightforward solution to gather the wireless sessions trace logs is to use already deployed Wi-Fi infrastructures. 
Several commercial alternatives are available in the market for the tracking of users using Wi-Fi infrastructure and the collection of the wireless sessions. Any of these alternatives allows us to gather data about the presence of users, and our problem requirements do not need an accurate location of the devices. The expected data gathered about the users is a log that stores and relates the user devices (the MAC address), the AP (the one that corresponds to the coverage area where the device is located in), and a timestamp. 

The output of this phase is the wireless session trace log which contains periodical data of all the devices and APs in the area under study. Each line of this trace log includes a tuple of three elements \texttt{<time\_stamp, user\_id, ap\_id>} stored periodically for all the devices and all the APs. By this, the wireless session of a given device can be generated and its handoffs can be also detected. From the point of view of a simulation of a fog infrastructure, a handoff is the only relevant event in the mobility of the users. A handoff involves that the service requests (or the data a device generates) come from a different AP and, probably, they also follow a different network route.

\subsection{Second phase: model definition}

Using the wireless session trace logs, the goal of this phase is to generate a mobility model using this log. In the context of our work, we propose to define the user mobility through the connection time of the devices to a given APs (represented with a time vector) and the probability of a change of the AP coverage area (represented with a stochastic transition matrix)~\cite{KERAMATJAHROMI2016137}. This type of model has been previously implemented in other scenarios such as the characterization of e-business workloads with Customer Behavior Model Graphs (CBMG)~\cite{10.5555/556438,10.1145/1071021.1071039}. 


A straightforward use of the transition matrices could result in mapping each entry of the matrix with one AP where devices are connected to, considering the transitions as the probabilities of a device handoff to another AP coverage area. This first approximation is not suitable for medium-big scenarios with a high number of APs in the infrastructure. In those cases, the complexity of the modeling is increased, resulting even in an unsolvable problem. Consequently, we propose to group APs, instead of mapping one AP to one state, and to divide the problem scenario into zones and levels with different granularity. 

For example, in the first level of a university campus, the region under study can be split into buildings, mapping each building with a zone (entry of the matrix) of the region under study. In successive iterations, the second levels (the buildings) are exploited by an isolated analysis of each of them, i.e., the modeling process is repeated for each building, which has to be split again into zones. 

By this hierarchical definition, the zones of the region are modeled independently and, consequently, the model can be modified by, for example, removing or duplicating zones. This modification simply results in the modification (removing/duplicating) of the rows and columns of the transition matrices and the time vectors. If the final user of the model is interested in applying the results to other different (but with similar features) mobility scenarios, the adaptation of an existing model is easier because of the possibility of modifying the zones of the model.





Figure~\ref{fig_stepsmobilitymodel} contains the steps we propose for the hierarchical modeling of the user mobility. In our hierarchical model, the analysis is split in successive levels of granularity to reduce the complexity, which results in the isolated repetition of some steps of this phase.

\subsubsection{Data cleaning}
Data cleaning refers to the analysis of the data set to identify and remove samples that are not useful for mobility modeling. It is a very particular process that depends on each data set and the target of the study. For example, a user can be interested in removing: users with only one AP sample; samples associated with a specific AP because it is located outside the region under study, samples associated with a specific device, samples in a time interval (during the night), etc.

\subsubsection{Hierarchical region definition}
Our proposal considers physical areas split into hierarchical zones. Each region is divided into zones, and the zones are recursively considered as sub-regions that are also divided into new zones. Each region corresponds to, or is modeled by, a pair of transition matrix and time vector. Similarly, the zones are related to the row and columns of the matrices/vectors.

Thus, we divide the characterization problem into levels in terms of geographic, topological, or other types of criteria. An independent modeling process needs to be carried out for each zone of a level recursively. This modeling process consists of the session identification, data filtering and mapping, clustering and transition matrix creation (Figure~\ref{fig_stepsmobilitymodel}).



\begin{figure}[!t]
\centering
\includegraphics[width=0.7\textwidth]{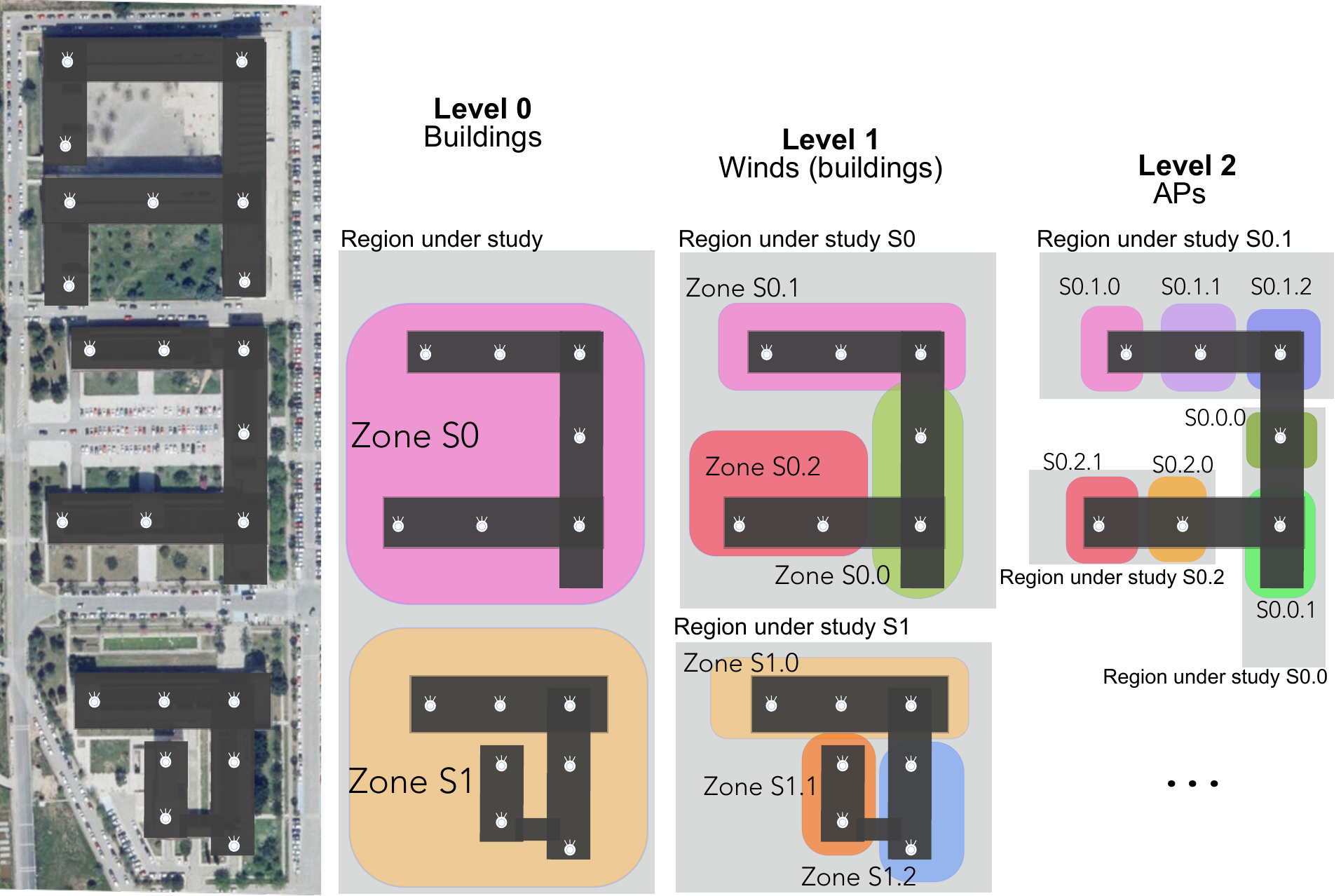}
\caption{Example of the hierarchical approach for mobility modeling.}
\label{fig_levelclustering}
\end{figure}

The hierarchy starts with the higher level, where the region under study is divided into $N$ zones, with $N$ being a reasonable number, neither too small (which reduces the model accuracy) nor too large (which increases the model complexity). Each zone is mapped to one state and they include disjoint and geographically adjacent APs. Once the modeling process is carried out for the first level, the process is repeated independently for each particular zone. The hierarchy is constrained by several requirements: the hierarchy levels are defined until the level with a granularity of 1 AP per zone is reached; each new region that emerges from a zone only considers APs into the former zone, the new zones covers all the APs (surjective) and APs are mapped with only one zone (disjoint)


We include an example of hierarchical division in Figure~\ref{fig_levelclustering}. The first level, level 0, divides the area under study into two zones which correspond to two buildings. The APs of each building are mapped to the corresponding zone. Once the model of level 0 is generated, the modeling process is repeated for each of the zones in level 0, i.e., a modeling process is executed for the first building (zone S0) and the second building (zone S1). These two new level 1 regions are split in terms of the wing of the buildings. This hierarchy is split in a reasonable number of levels until each zone only covers one AP, as in our example which requires three levels.

The use of a hierarchical division of the region under study also facilitates the generalization of the resulting model. As we previously explained, the number of open mobility models is very reduced and the possibility to generalize the existing one is an important point. A hierarchical model easily allows to remove, duplicate, or modify some regions of the resulting model to adapt it to a different study case. Consider an open model that considers three buildings of a university campus, one building is a student housing, another one contains the teaching classrooms, and the teachers' office rooms are in the last one. If we need to adapt this model to a different campus, for example with three classroom buildings or without student housings, we can adapt it by duplicating the part corresponding to the classroom building or by removing the model of the housing building. This generalization can be also performed in terms of user types, popularity probabilities, etc.

\subsubsection{Device session identification} \label{proposed_sol_device_identification}
Once the data is loaded and cleaned, the next step is to identify the users. We assume that a MAC address corresponds to one user. Consequently, the selection of the samples of a user requires to filter by the MAC address. Not only the users need to be identified, but also the user sessions. A session is defined as each of the \textit{visits} that a user does to the region under study during the time of the experiment. In other words, if a user leaves the region under study and comes back again, these two \textit{visits} result in two different user sessions. 

From the point of view of the infrastructure, a user session is defined as a device that enters the coverage perimeter defined by the levels of the study at a certain point in time, it is in the area for a specific period of time, and it leaves it. The Wi-Fi infrastructure does not register the disconnections in the data sets and, consequently, the session detection is conceptually determined when:
\begin{itemize}
    \item The user changes between regions under study.
    \item None connections are detected after a certain period of time.
\end{itemize}

And these two criteria are carried out over the data set by detecting when:

\begin{itemize}
    \item A sample of a user is associated with an AP that does not belong to any of the zones of the current area under study.
    
    \item The time between two successive connections of the same device is greater than a defined threshold value.
\end{itemize}

The definition of the threshold value for the user session identification implies carrying out a preliminary study to calculate the optimal threshold time for the data set that is being analyzed. For example, in our case study in Section~\ref{sect:casestudy}, a range of defined threshold values are tested and, for all of them, the Euclidean distance between the number of sessions and the average connection time of these is calculated. The optimal threshold is determined by the smaller Euclidean distance, based on the idea that the best threshold is the one that balances the total number of sessions and the length of the sessions~\cite{10.1145/1071021.1071039}.

\subsubsection{Data filter / map}
Once the different user sessions are identified, not all their samples in the data set provide useful information. Consequently, the samples are filtered by keeping:
\begin{itemize}
    \item The first sample of a consecutive subset of samples with the same AP identifier. This corresponds to the handoffs (AP changes) of a user session.
    \item The first and the last sample of each session. They identify the entry and exit points of the user session.
\end{itemize}

The result of this phase is a subset of samples that represents the session start, end and the AP changes. Figure~\ref{fig_usertrack}.c shows an example of the subset obtained from Figure~\ref{fig_usertrack}.b.

\subsubsection{User profiling: clustering}
\label{sect_modelclustering}

Human behavior is very different and it cannot be determined only with a general model, and the same pattern for all users is not realistic nor accurate. Consequently, it is important to split users with similar mobility patterns. The goal of user profiling or segmentation is to achieve a clear description of how the users in each cluster behave and move around the area under study. Clustering is the usual solution for this type of profiling and it is necessary to trade off the number of user types (as smaller as possible) and the suitability of the model (as accurate as possible)~\cite{HARDY199683}.

We characterize the behavior of the users in terms of the percentage of time that they spend in a specific zone of the region under study. Consequently, each user session is characterized with a n-tuple probability vector, where $n$ is the number of zones in the region under study. This vector represents the probability of the user being in each of the zones, calculated as the total time in each zone divided by the session length.

The selection of a clustering algorithm depends on many factors~\cite{abbas2008comparisons}. For example, in our case study, we implement the k-means algorithm because of its simplicity and its generalized use.

\subsubsection{Transition matrix and time vector creation}

Once the users are split into different groups according to their behavior using a clustering process, a data model is generated for each of the clusters. This data model is made up of a transition matrix and a vector with the average stay length in each zone.

The transition matrix represents the transitions between zones, where each element $a_{i,j}$ of the matrix is the probability to go to a zone $j$ when a user is in zone $i$. Each element $a_{i,j}$ is calculated as the total number of handoffs from $i$ to $j$ divided by the total number of handoffs for all the sessions of the user in a given cluster group. Additionally, the matrix is completed with two additional elements labeled as \texttt{IN} and \texttt{OUT}, that respectively represent the start and the end of a user session. In other words, $a_{IN,j}$ represents the probability that the zone $j$ is the starting point of a user. On the contrary, $a_{i,OUT}$ represents the probability that the zone $i$ is the last zone of a user session. Consequently, $a_{OUT,j}=0\ \forall j$ and  $a_{i,IN}=0\ \forall i$. 

The stay length vector indicates the average time that a given user type spends in each zone expressed in time units. Each element of the vector $v_i$ is calculated as the sum of the time of all the visits of a user to a zone $i$ divided by the total number of visits to that zone. 


The data models provide a double utility. First, they serve to characterize the movement of the users of the different groups. They also are, if necessary, the input for the next phase, the synthetic generation of traces.

\subsection{Third phase: generation of the synthetic traces}

The objective of this phase is to generate a synthetic data set using the data model obtained in the previous phase. The most important advantage of the generation of synthetic traces is that the resulting model can be extrapolated to scenarios with other features (for example, changing the number of users, building, removing some user type, etc.). Additionally, the use of synthetic traces also guarantees additional anonymization of sensitive information.


The synthetic data is also a three elements tuple: \texttt{<time\_stamp, user\_id, ap\_id>}, in front of the original \texttt{<time\_stamp, device\_id, ap\_id>} tuple. The user identifier for synthetic tuples is  auto-incrementally generated during the synthetic generation.

Each user trace is generated randomly using the mobility model. Each time synthetic users are generated, we first randomly determine the user types they belong to, using either weights determined manually or by the popularity distribution of the real data set. Using the probabilities of the transition matrix of that user group, the entry point (zone) is determined. The time in that zone is determined with the stay period vector and randomly generated following an exponential distribution. The second zone in the user trace is again determined with the probabilities of the transition matrix. This is repeated subsequently until the zone \textit{OUT} is chosen randomly.

%
%

The times of each sample, \texttt{<time\_stamp, device\_id, ap\_id>}, are mapped to the simulation time, which starts in time 0 and are incremented in simulation time units. The generation of the users is determined by increasing the time of their initial sample with the average time between user arrivals. This time can be manually fixed or we can use the one obtained from the real data set. An exponential distribution is used for the generation of these user arrivals.

\section{Case study: the campus of the \uiblong}
\label{sect:casestudy}

We apply our hierarchical decomposition method in the campus of the \uiblong\  (\uibacro). \uibacro\ is a small-medium size university with approximately 14,000 students, 1,000 teachers and 800 administrative staff. The Wi-Fi infrastructure of the university is composed of 425 APs, distributed along with the 18 buildings in the campus, which are located in a secluded area of the city, specific for the campus.

In the remainder of this section, we detail the specific aspects to apply the method proposed in Section~\ref{sect_proposedsolution} to the case of the \uibacro. At the end of the section, some of the results obtained for the mobility model are also commented.

\subsection{Data collection}
\label{sect_datacollection}

The APs deployed in the \uibacro\ are from the company Aruba Networks. These APs include the ALE technology (Analytics and Location Engine) which collects location and mobility data of the user devices~\cite{10.1145/3342428.3342652}. Consequently, we easily implement the data collection phase by using the methods provided by ALE.

We first use the method \texttt{access\_points} to collect the list and data of all the APs deployed in the Wi-Fi infrastructure of the campus. Secondly, ALE offers data about the location of the devices/users with three main methods: \texttt{proximity}, \texttt{presence} and \texttt{station}. We choose \texttt{proximity}, since it registers the MAC addresses of all the detected user devices, whether or not they are logged to the Wi-Fi network. Method \textit{proximity} also associates the user device with the closest AP. Figure~\ref{fig:logale} shows an example of the JSON file returned by the method  \texttt{proximity}. In this example, data about only one device and one AP are represented, but a usual execution returns all the devices for all the APs.

\begin{figure}
\begin{Verbatim}[fontfamily=zi4, numbers=left, numbersep=5pt, fontsize=\scriptsize, numberblanklines=false, 
frame=single, framesep=1mm, framerule=0.1pt, rulecolor=\color{gray},  firstnumber=1,tabsize=2]
{  "Proximity_result": [{
    "msg": {
      "sta_eth_mac": {
        "addr": "FCF8AEEC8E20"
      },
      "associated": true,
      "hashed_sta_eth_mac": "DBB3D96B0AA540A2839B39071C3154760775AD50"
      "ap_name": "9c:1c:12:c0:19:5a"
      "radio_mac": {
        "addr": "B45D50F94E30"
      },
      "target_type": "TARGET_TYPE_STATION"
    },
    "ts": 1521133688,
}]}
\end{Verbatim}
\caption{Example of device tracking log obtained with a commercial tool (Aruba Analytics and Location Engine).}
\label{fig:logale}
\end{figure}




Note that our API vendor only offers real-time data. Users interested in analyzing historical data need to store the data persistently, for example, in a database. We implemented a Python script which requests the method \texttt{proximity}, parses the returned JSON file, and stores into a database a 3-tuple \texttt{<time\_stamp, user\_id, ap\_id>} for each device in the coverage areas.  ALE is configured to return the one-way hashed MAC addresses to protect the privacy of the users. The hash function is \textit{salted} for periods of one day to avoid the matching of traces with real users.  The script is executed each minute, so the database stores the device-AP connections with an accuracy of 1 minute. The database stores the hashed MAC address of the user device as the device identifier, and the MAC address of the AP as the AP identifier.

For this study, we gathered the Wi-Fi probes with a minute frequency during one week, November 9th-15th 2020, obtaining 122 MB of raw data. In this period, partial mobility restrictions were applied due to COVID-19 pandemic. The presence was reduced to approximately half of the usual number of users.
Additionally, we conduct several studies, by dividing the data of this week in different time subsets. More concretely, 9 studies are considered: one for each single day of the week, one for the five working days, and one for the weekend. 
We chose one of them, the experiment of Monday 9th November, to be analyzed in the results of this paper. The selected experiment is a suitable example for any of the working days. If the reader is interested in the other results, they can be accessed in the data repository associated with this work~\footnote{\url{\github}}.

\subsection{Model definition}

\subsubsection{Data cleaning}


In the data cleaning process we remove the samples of sessions formed by a single connection. Thus we avoid samples of passersby who simply pass near the campus but do not make any route within it.

\subsubsection{Region definition}

Using the hierarchical definition of our method, we define three levels in the division of the university campus. At the initial level (level 0), the entire campus is considered, dividing it into as many zones as buildings. Table~\ref{tab:zoneslevel0} shows the 18 zones, with a range of APs per zone between 1 and 63.

At the second level (level 1), we study  the movements individually for each building. We split each building into floors and wings. Finally, at the last level (level 2), each floor-wing is divided into zones that only includes one AP.


Each region under study has a 3-tuple identifier based on the level, the building acronym and the floor-wing, $\left< level, building, wing \right>$. For example, $\left<level\_0, -, - \right>$ refers to the analysis of level 0 that corresponds to the movement of users between buildings, $\left< level\_1, buildingA, - \right>$ refers to the analysis of mobility of the users between wings of \textit{Building A}, or $\left< level\_2, buildingA, northSecondFloor \right>$ refers to the analysis of the user movement between the APs in the north wing of the second floor of \textit{Building A}.


The high number of regions by level inhibits a detailed description of each of them in this article. For this reason, we only explain one region from each level. More concretely, the explained regions under study are  \cslzero, \cslone, and \csltwo, all of them from the period corresponding to Monday 9th November 2020. The results for the other regions under study are published in the data repository~\footnote{\url{\github}}.

\begin{table}[h]
    \centering
    \begin{tabular}{lc|lc}
      \hline
    \textit{Zone id.} & \textit{Number of APs} & \textit{Zone id.} & \textit{Number of APs} \\ \hline
    
bldg\_AT $^{1}$& 36&bldg\_JO& 63\\  \hline
bldg\_CEP& 14&bldg\_MA& 43\\  \hline
bldg\_CJ& 7&bldg\_MEN& 8\\  \hline
bldg\_CL& 1&bldg\_MO& 52\\  \hline
bldg\_CTI& 9&bldg\_RES& 18\\  \hline
bldg\_EIV& 11&bldg\_RL& 42\\  \hline
bldg\_GC& 35&bldg\_SCT& 27\\  \hline
bldg\_IE& 7&bldg\_SE& 8\\  \hline
bldg\_ITD& 16&bldg\_SL& 28  \\  \hline
    \end{tabular}
\begin{flushleft}
\footnotesize
    $^{1}$ Zone selected as the region under study for level 1.
\end{flushleft}
    \caption{Division into zones of the region under study \cslzero}
    \label{tab:zoneslevel0}
\end{table}

\begin{table}[h]
    \centering
    \begin{tabular}{lc|lc}
      \hline
    \textit{Zone id.} & \textit{Number of APs} & \textit{Zone id.} & \textit{Number of APs} \\ \hline
    
basement\_fl&3& 1st\_fl\_East&6 \\  \hline
0\_fl\_North & 7 & 1st\_fl\_North  & 7 \\  \hline
0\_fl\_East $^{1}$&6& 2nd\_fl &3\\  \hline
    \end{tabular}
\begin{flushleft}
\footnotesize
    $^{1}$ Zone selected as the region under study for level 2.
\end{flushleft}
    \caption{Division into zones of the region under study \cslone}
    \label{tab:zoneslevel1}
\end{table}

Table~\ref{tab:zoneslevel0} shows the number of APs in each zone of the region under study \cslzero. At this level, one zone is created for each of the buildings on the campus.

Table~\ref{tab:zoneslevel1} shows the number of APs in each of the zones defined in the building $bldg\_AT$, used as an example of the region under study for level 1. It is a four-storey building and two wings (north and east), with an approximate occupation of 800 students and 100 teachers where, due to the architecture of the building, we delimit one single zone for the basement and the second floor, and two zones --one for each wing-- for the ground floor and the first floor.

Finally, the wing $0\_fl\_East$ is selected as an example of the region under the study of level 2. This is the deepest level in the region hierarchy and each APs is defined as an isolated zone. Thus, 6 zones are considered, one for each AP.

\begin{figure}[h!]
    \centering
    \includegraphics[width=0.7\textwidth]{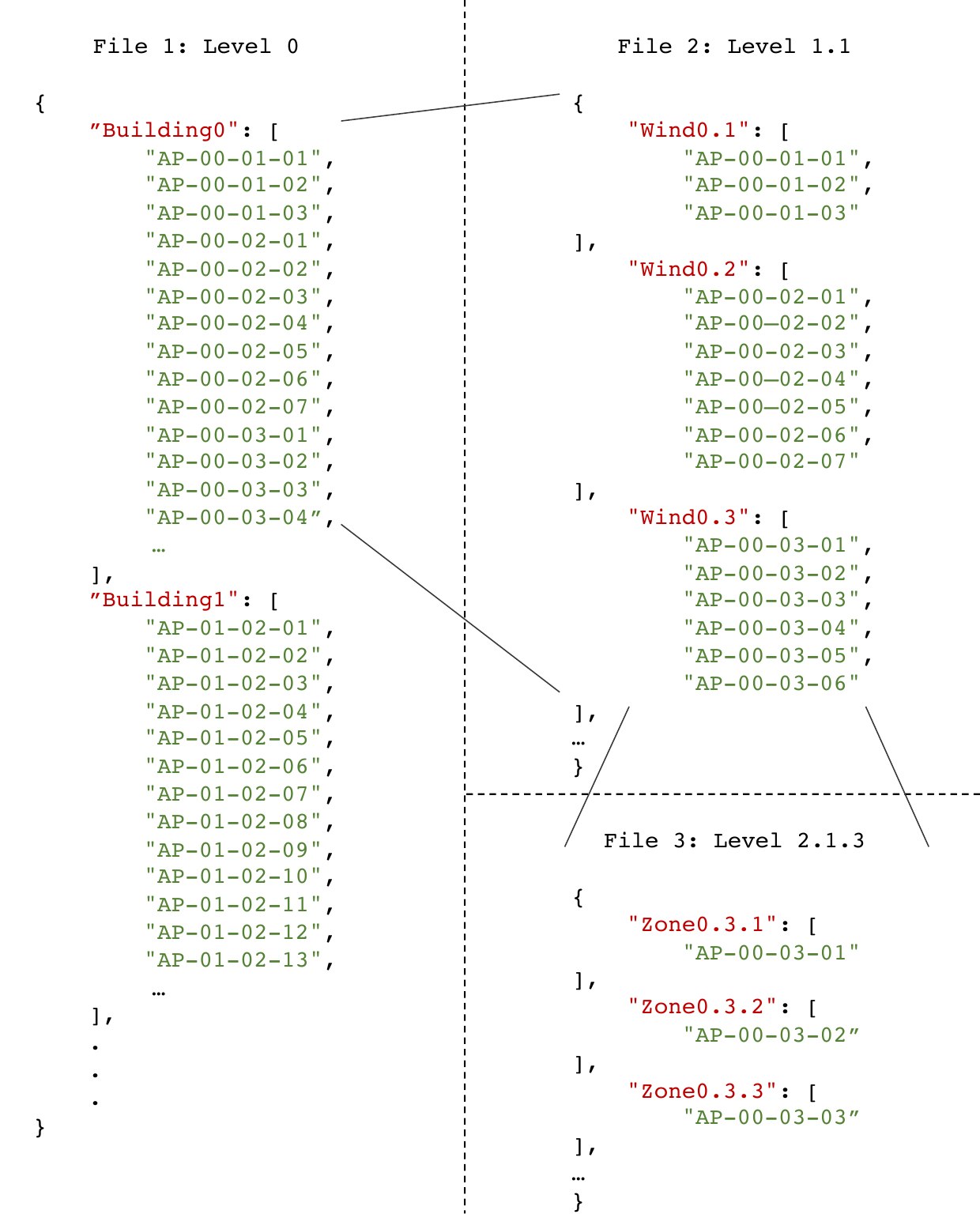}
    \caption{Example of three JSON files for the hierarchical level definition.}
    \label{fig:json_lvls}
\end{figure}

We implement the definition of each of the levels using a JSON file that specifies the APs included in each zone. Figure~\ref{fig:json_lvls} shows the source code of the JSON files corresponding to the three areas under study explained in this section.

\subsubsection{Device session identification}

As we commented in Section~\ref{proposed_sol_device_identification}, the user disconnections are not registered in the trace logs and user sessions may be determined manually. An iterative process, with an incremental threshold value, calculates the number of sessions and the average connection time over all the users for each threshold value. The desired threshold is the one that reduces both the number of sessions and the average session length. Consequently, we use the Euclidean distance because it indicates the threshold value which balances both metrics. The threshold value depends on each data set and, consequently, the process of threshold definition must be repeated for each region under study.

\begin{figure}[!h]
    \centering
    \subfigure[Threshold calculation for \cslzero]{
        \includegraphics[width=0.45\textwidth, page=1]{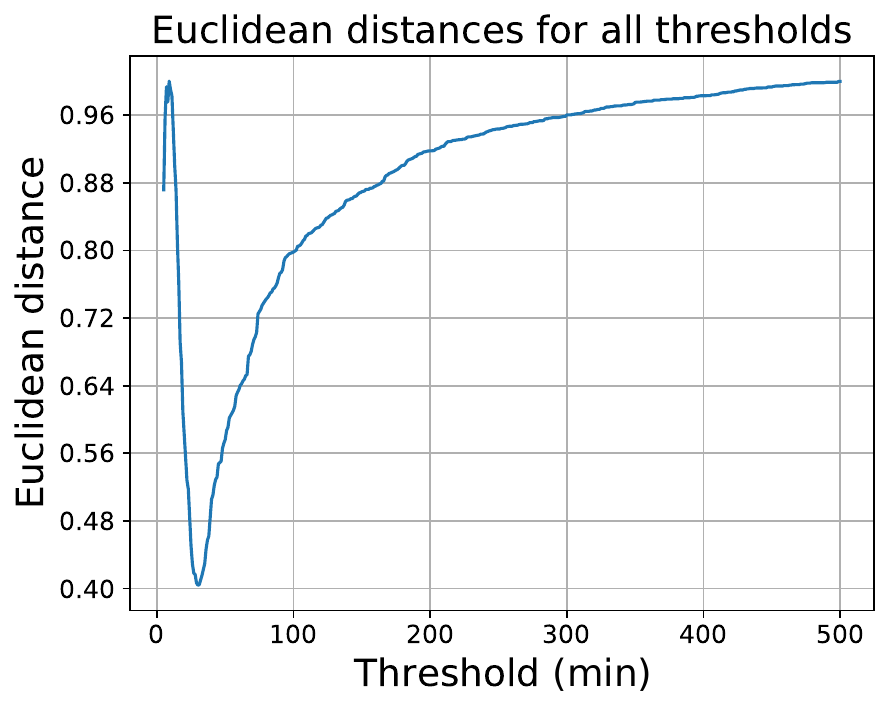}
        \includegraphics[width=0.45\textwidth, page=2]{figuresX/threshold_lvl0.pdf}
        }
    \subfigure[Threshold calculation for \cslone]{
        \includegraphics[width=0.45\textwidth, page=1]{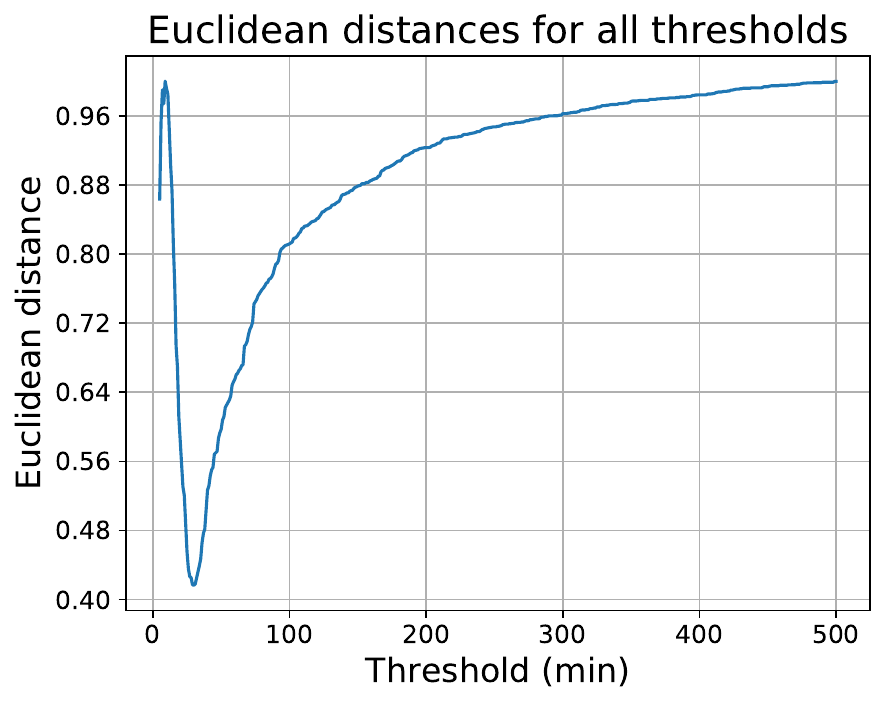}
        \includegraphics[width=0.45\textwidth, page=2]{figuresX/threshold_lvl1.pdf}
        }
    \subfigure[Threshold calculation for \csltwo]{
        \includegraphics[width=0.45\textwidth, page=1]{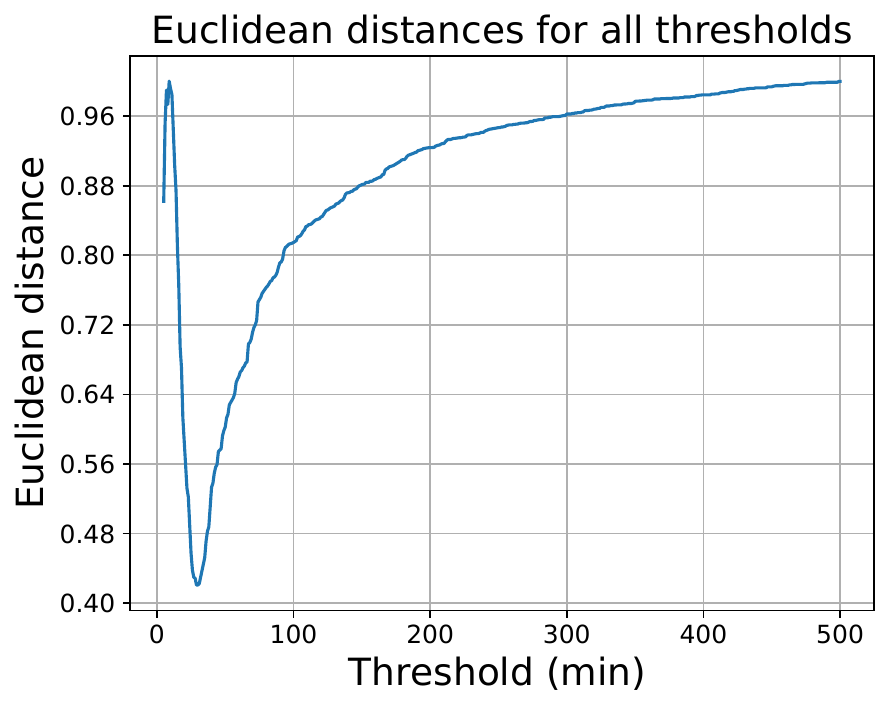}
        \includegraphics[width=0.45\textwidth, page=2]{figuresX/threshold_lvl2.pdf}
        }

    \caption{Euclidean distances between number of user sessions and average session times.}
    \label{fig:threshold}
\end{figure}



Figure~\ref{fig:threshold} represents the evolution of the Euclidean distance as the threshold value is increased. Three plots are presented, one for each region included in this section. 


For \cslzero, the smaller Euclidean distance results in a threshold of 30 minutes. This threshold corresponds to the case of 11,520 user sessions with 4,883.7 seconds of average session time. In the example for level 1, \cslone, the threshold is also defined in 30 minutes, with 11,570 user sessions and 4,737.0 seconds of average time. Finally, \csltwo\ results in a threshold of 29 minutes with 12,599 users and 4,252.7 seconds of session time.

Once the threshold value is selected, the user sessions can be recalculated considering the criteria explained in Section~\ref{proposed_sol_device_identification}.



\subsubsection{Data filter / map}

The data filtering is straightforward, and it only requires to loop through each user session, taking into account the threshold value of the previous phase. The process keeps the first sample in the session, the first sample of a new AP subset, and the last sample in the user session. All the other samples are removed.

\subsubsection{User profiling: clustering}

In our case study, we implement the k-means algorithm for the clustering of the user sessions. Note that the user mobility is characterized by their zone probability vector, the vector that indicates the probability for a given user of being in a given zone of the region under study. Consequently, we first calculate those vectors for each user session. 

\begin{figure}[!h]
    \centering
    \subfigure[\cslzero]{
        \includegraphics[width=0.45\textwidth, page=1]{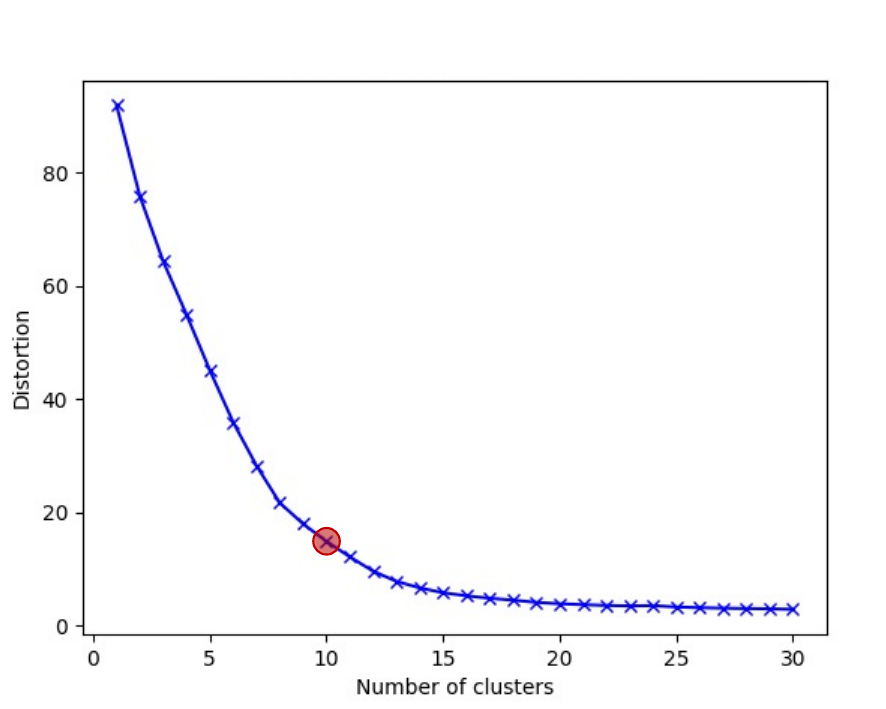}
        }
    \subfigure[\cslone]{
        \includegraphics[width=0.45\textwidth, page=1]{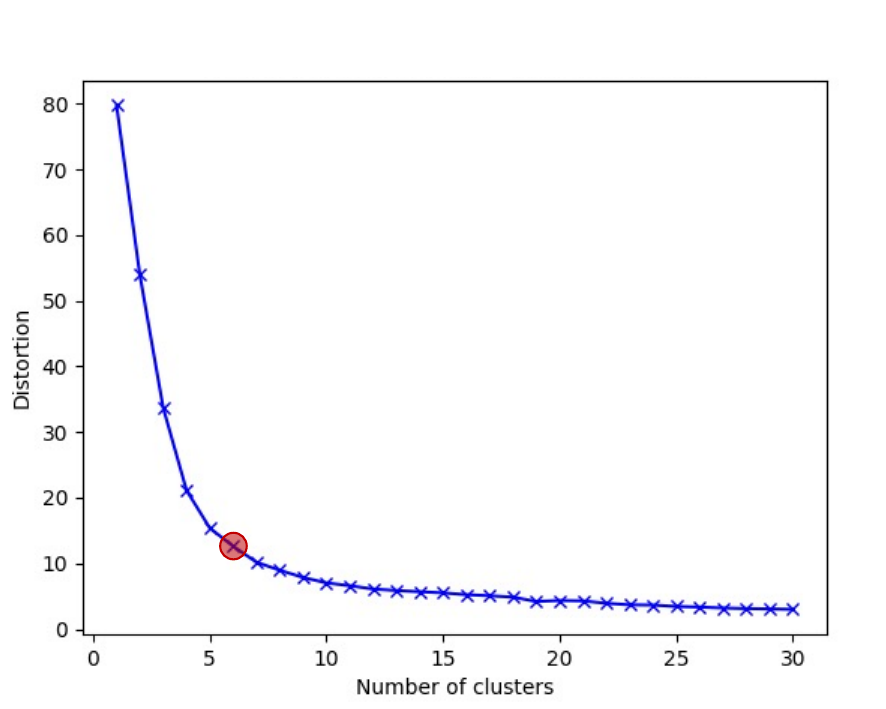}
        }
    \subfigure[\csltwo]{
        \includegraphics[width=0.45\textwidth, page=1]{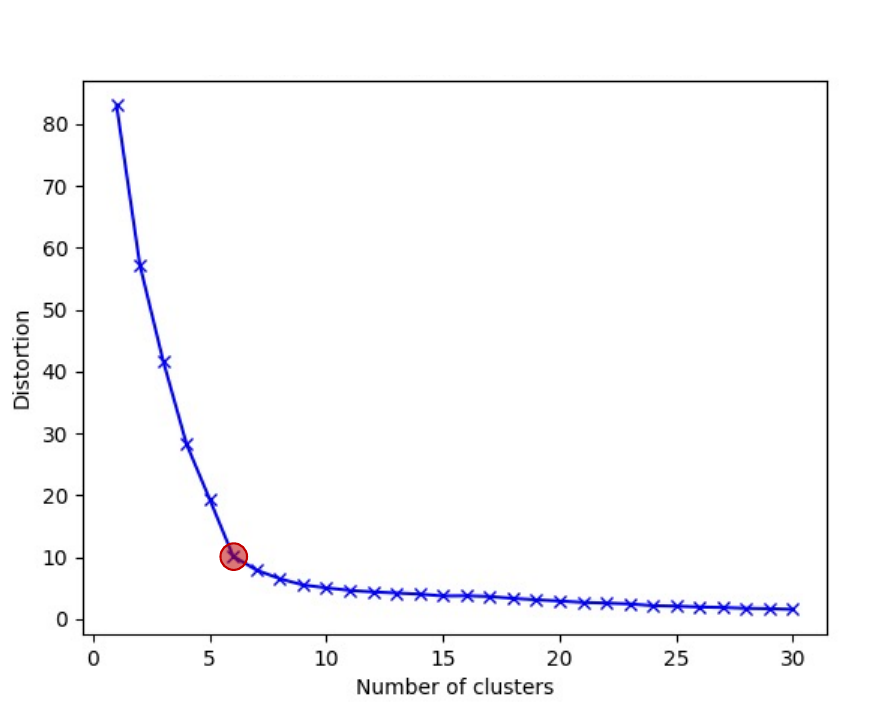}
        }
    \caption{Elbow method to detect the optimal number of clusters considering the distortion metric and with a maximum number of 30 clusters.}
    \label{fig:elbow}
\end{figure}

In the case of the k-means algorithm, similarly to some other clustering methods, the number of clusters is an input value of the algorithm. Thus, a previous study of the suitable number of clusters needs to be performed. We calculate it through the elbow method~\cite{elbow_cluster}. The objective behind this is to minimize both opposite metrics of intergroup distance\footnote{We measure the intergroup distance using the distortion metric. The distortion is calculated as the average of the squared distances from the cluster centers of the respective clusters.} and the number of clusters. The elbow method determines the optimal number of clusters $k$ like the number of clusters from which the improvement, in terms of intergroup distance, obtained by increasing this number of clusters is so small that it does not justify the additional cost (complexity) of increasing the number of clusters. This is graphically determined in the \textit{elbow} of the curve, hence the name of the method (Figure~\ref{fig:elbow}).

Figure~\ref{fig:elbow} shows the \textit{elbows} for the three regions under study that we present in this section. We can see that the optimal number of clusters is 10 for \cslzero , 6 for \cslone , and 6 for \csltwo. Once the optimal number of clusters is defined, the k-mean algorithm is executed and the clusters for each region are generated. 






\subsubsection{Transition matrix and time vector creation}

The final step is the generation of the transition matrices and the time vectors. In this step, we obtain one pair of matrix-vector for each user cluster, i.e., ten matrices and vectors for \cslzero, six for \cslone, and six for \csltwo, because the total numbers of user profiles (clusters) for each case are 10, 6, and 6. 

Instead of representing these 22 matrices/vectors, we only show one pair of matrix-vector from each of the three regions. The other ones are available in the open repository. Equation~\ref{eq:mobility_case9} shows the transition matrix for one of the 10 user profiles generated in \cslzero, and Equation~\ref{eq:times_case9} the average stay time in minutes for each building of that region. Equation~\ref{eq:mobility_l1} and Equation~\ref{eq:time_l1} respectively show the transition matrix and the time vector of one of the 6 user profiles in \cslone. Finally, Equation~\ref{eq:mobility_l2} and Equation~\ref{eq:time_l2} correspond to the mobility model of one of the six profiles in \csltwo.


\begin{equation}\label{eq:mobility_case9}
\resizebox{0.9\hsize}{!}{%
  \begin{blockarray}{cccccccccccccccc}
 &  \BAmulticolumn{14}{c}{\Large TM\cslzero = } & \\
 \\
     & \textcolor{gray}{IN} & \textcolor{gray}{OUT} & \textcolor{gray}{bldg\_AT} & \textcolor{gray}{bldg\_CEP} & \textcolor{gray}{bldg\_CJ} & \textcolor{gray}{bldg\_GC} & \textcolor{gray}{bldg\_IE} & \textcolor{gray}{bldg\_ITD} & \textcolor{gray}{bldg\_JO} & \textcolor{gray}{bldg\_MA} & \textcolor{gray}{bldg\_MO} & \textcolor{gray}{bldg\_RES} & \textcolor{gray}{bldg\_RL} & \textcolor{gray}{bldg\_SCT} & \textcolor{gray}{bldg\_SL}\\ 
	  \begin{block}{r[ccccccccccccccc]}
	  	\textcolor{gray}{IN} & 0 & 0 & 0.0033 & 0 & 0 & 0.0033 & 0.0042 & 0 & 0.0025 & 0.0017 & 0.0042 & 0.9791 & 0 & 0 & 0.0017\\ 
			\textcolor{gray}{OUT} & 0 & 0 & 0 & 0 & 0 & 0 & 0 & 0 & 0 & 0 & 0 & 0 & 0 & 0 & 0\\
			\textcolor{gray}{bldg\_AT} & 0 & 0.364 & 0 & 0 & 0 & 0 & 0 & 0.091 & 0 & 0.273 & 0 & 0.273 & 0 & 0 & 0\\ 
			\textcolor{gray}{bldg\_CEP} & 0 & 1 & 0 & 0 & 0 & 0 & 0 & 0 & 0 & 0 & 0 & 0 & 0 & 0 & 0\\ 
			\textcolor{gray}{bldg\_CJ} & 0 & 0.5 & 0 & 0 & 0 & 0 & 0 & 0 & 0 & 0 & 0.5 & 0 & 0 & 0 & 0\\ 
			\textcolor{gray}{bldg\_GC} & 0 & 0.333 & 0 & 0 & 0 & 0 & 0 & 0 & 0 & 0 & 0.0667 & 0.4 & 0.0667 & 0.0667 & 0.0667\\ 
			\textcolor{gray}{bldg\_IE} & 0 & 0.2069 & 0 & 0 & 0.0345 & 0 & 0 & 0 & 0.069 & 0 & 0 & 0.69 & 0 & 0 & 0\\ 
		\textcolor{gray}{bldg\_ITD} & 0 & 0.5 & 0 & 0 & 0 & 0.25 & 0 & 0 & 0 & 0 & 0 & 0.25 & 0 & 0 & 0\\

		\textcolor{gray}{bldg\_JO} & 0 & 0.364 & 0 & 0 & 0 & 0 & 0 & 0 & 0 & 0.0909 & 0 & 0.545 & 0 & 0 & 0\\
		\textcolor{gray}{bldg\_MA} & 0 & 0.182 & 0.273 & 0 & 0 & 0 & 0.091 & 0 & 0.091 & 0 & 0 & 0.364 & 0 & 0 & 0\\
		\textcolor{gray}{bldg\_MO} & 0 & 0.5 & 0 & 0 & 0 & 0.143 & 0 & 0 & 0 & 0 & 0 & 0.286 & 0.0714 & 0 & 0\\
		\textcolor{gray}{bldg\_RES} & 0 & 0.952 & 0.0033 & 025 & 0.0008 & 0.0057 & 0.019 & 0.0025 & 0.0041 & 0.0041 & 0.0041 & 0 & 0.0025 & 0 & 0\\ 
		\textcolor{gray}{bldg\_RL} & 0 & 0.2 & 0 & 0 & 0 & 0.2 & 0 & 0 & 0 & 0 & 0.2 & 0.4 & 0 & 0 & 0\\ 
		\textcolor{gray}{bldg\_SCT} & 0 & 0 & 0 & 0 & 0 & 0 & 0 & 0 & 0 & 0 & 1 & 0 & 0 & 0 & 0\\ 
		\textcolor{gray}{bldg\_SL} & 0 & 0 & 0 & 0 & 0 & 0 & 0 & 0 & 0 & 0 & 0 & 1 & 0 & 0 & 0 \\
	  \end{block}
  \end{blockarray}
}
\end{equation}

\begin{equation}\label{eq:times_case9}
\resizebox{0.9\hsize}{!}{%
  \begin{blockarray}{cccccccccccccc}
   &  \BAmulticolumn{12}{c}{ \large TV\cslzero = } & \\
\\
     & \textcolor{gray}{bldg\_AT} & \textcolor{gray}{bldg\_CEP} & \textcolor{gray}{bldg\_CJ} & \textcolor{gray}{bldg\_GC} & \textcolor{gray}{bldg\_IE} & \textcolor{gray}{bldg\_ITD} & \textcolor{gray}{bldg\_JO} & \textcolor{gray}{bldg\_MA} & \textcolor{gray}{bldg\_MO} & \textcolor{gray}{bldg\_RES} & \textcolor{gray}{bldg\_RL} & \textcolor{gray}{bldg\_SCT} & \textcolor{gray}{bldg\_SL}\\
	  \begin{block}{c[ccccccccccccc]}
			& 53.32 & 40.48 & 1.01 & 27.46 & 14.07 & 5.62 & 17.2 & 28.5 & 19.89 & 89.59 & 77.22 & 9.08 & 9.17\\
	  \end{block}
  \end{blockarray}
}
\end{equation}

\begin{equation}\label{eq:mobility_l1}
\resizebox{0.7\hsize}{!}{%
 		\begin{blockarray}{cccccccc}
 		   &  \BAmulticolumn{6}{c}{  TM\cslone = } & \\
\\
		& \textcolor{gray}{IN} & \textcolor{gray}{OUT} & \textcolor{gray}{1st\_fl\_East} & \textcolor{gray}{1st\_fl\_North} & \textcolor{gray}{0\_fl\_East} & \textcolor{gray}{0\_fl\_North} & \textcolor{gray}{2nd\_fl}\\
		\begin{block}{r[ccccccc]}
			\textcolor{gray}{IN} & 0 & 0 & 0.0364 & 0.0182 & 0.0364 & 0 & 0.909\\
			\textcolor{gray}{OUT} & 0 & 0 & 0 & 0 & 0 & 0 & 0\\
			\textcolor{gray}{1st\_fl\_East} & 0 & 0.4 & 0 & 0 & 0 & 0 & 0.6\\
			\textcolor{gray}{1st\_fl\_North} & 0 & 0.286 & 0 & 0 & 0 & 0 & 0.714\\
			\textcolor{gray}{0\_fl\_East} & 0 & 0.25 & 0 & 0 & 0 & 0 & 0.75\\
			\textcolor{gray}{0\_fl\_North} & 0 & 0.5 & 0 & 0 & 0 & 0 & 0.5\\
			\textcolor{gray}{2nd\_fl} & 0 & 0.568 & 0.243 & 0.081 & 0.081 & 0.027 & 0\\
		\end{block} 
	\end{blockarray}
}
\end{equation}
	
\begin{equation}\label{eq:time_l1}
\resizebox{0.5\hsize}{!}{%
\begin{blockarray}{cccccc}
   &  \BAmulticolumn{4}{c}{  TV\cslone = } & \\
\\
& \textcolor{gray}{1st\_fl\_East} & \textcolor{gray}{P1N} & \textcolor{gray}{0\_fl\_East} & \textcolor{gray}{0\_fl\_North} & \textcolor{gray}{2nd\_fl}\\
\begin{block}{c[ccccc]}
 & 6.92 & 15.03 & 14.93 & 6.23 & 51.85 \\
\end{block}
\end{blockarray}
}
\end{equation}

\begin{equation}\label{eq:mobility_l2}
\resizebox{0.7\hsize}{!}{%
	  \begin{blockarray}{ccccccccc}
	   		   &  \BAmulticolumn{7}{c}{  TM\csltwo = } & \\
\\
	     & \textcolor{gray}{IN} & \textcolor{gray}{OUT} & \textcolor{gray}{AP-OO-03-01} & \textcolor{gray}{AP-OO-03-02} & \textcolor{gray}{AP-OO-03-03} & \textcolor{gray}{AP-OO-03-04} & \textcolor{gray}{AP-OO-03-05} & \textcolor{gray}{AP-OO-03-06}\\
	  \begin{block}{c[cccccccc]}
			\textcolor{gray}{IN} & 0 & 0 & 0.0058 & 0.953 & 0.035 & 058 & 0 & 0\\
			\textcolor{gray}{OUT} & 0 & 0 & 0 & 0 & 0 & 0 & 0 & 0\\
			\textcolor{gray}{AP-OO-03-01} & 0 & 0.75 & 0 & 0 & 0.25 & 0 & 0 & 0\\
			\textcolor{gray}{AP-OO-03-02} & 0 & 0.825 & 0.0164 & 0 & 0.104 & 0.027 & 0.016 & 0.011\\
			\textcolor{gray}{AP-OO-03-03} & 0 & 0.333 & 0 & 0.667 & 0 & 0 & 0 & 0\\
			\textcolor{gray}{AP-OO-03-04} & 0 & 0.5 & 0 & 0.333 & 0.167 & 0 & 0 & 0\\
			\textcolor{gray}{AP-OO-03-05} & 0 & 1 & 0 & 0 & 0 & 0 & 0 & 0\\
			\textcolor{gray}{AP-OO-03-06} & 0 & 1 & 0 & 0 & 0 & 0 & 0 & 0\\
	  \end{block}
	  \end{blockarray}
}
\end{equation}

\begin{equation}\label{eq:time_l2}
	\resizebox{0.5\hsize}{!}{%
	 \begin{blockarray}{ccccccc}
	    &  \BAmulticolumn{5}{c}{  TV\csltwo = } & \\
\\
    	 & \textcolor{gray}{AP-OO-03-01} & \textcolor{gray}{AP-OO-03-02} & \textcolor{gray}{AP-OO-03-03} & \textcolor{gray}{AP-OO-03-04} & \textcolor{gray}{AP-OO-03-05} & \textcolor{gray}{AP-OO-03-06}\\
	  \begin{block}{c[cccccc]}
		t = & 4.6 & 32.4 & 6.35 & 8.58 & 0 & 0\\
	  \end{block}
	  \end{blockarray}
}
\end{equation}

\begin{figure}[!h]
    \centering
    \subfigure[\cslzero]{
        \includegraphics[width=0.44\textwidth, page=1]{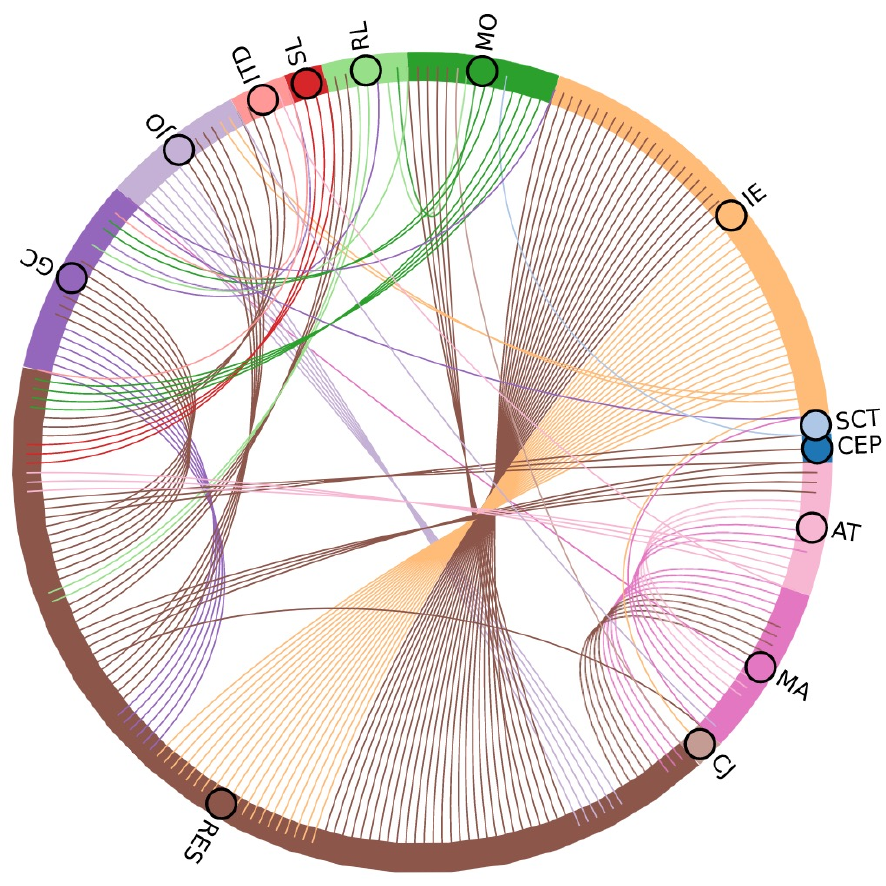}
        }
    \subfigure[\cslone]{
        \includegraphics[width=0.45\textwidth, page=1]{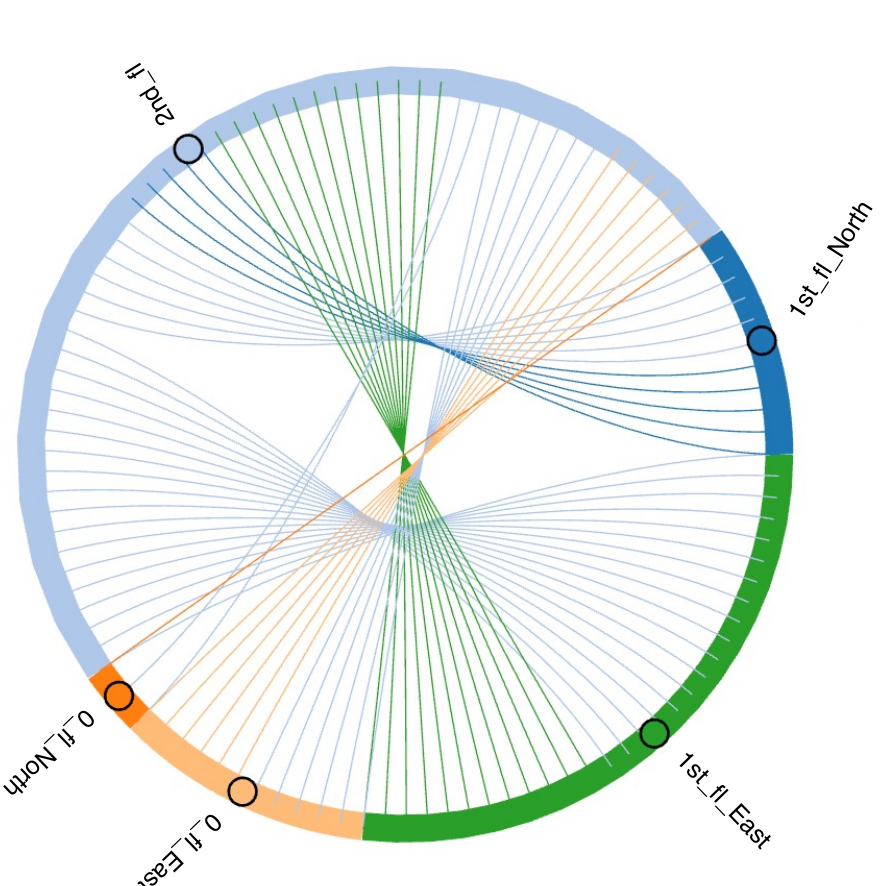}
        }
    \subfigure[\csltwo]{
        \includegraphics[width=0.55\textwidth, page=1]{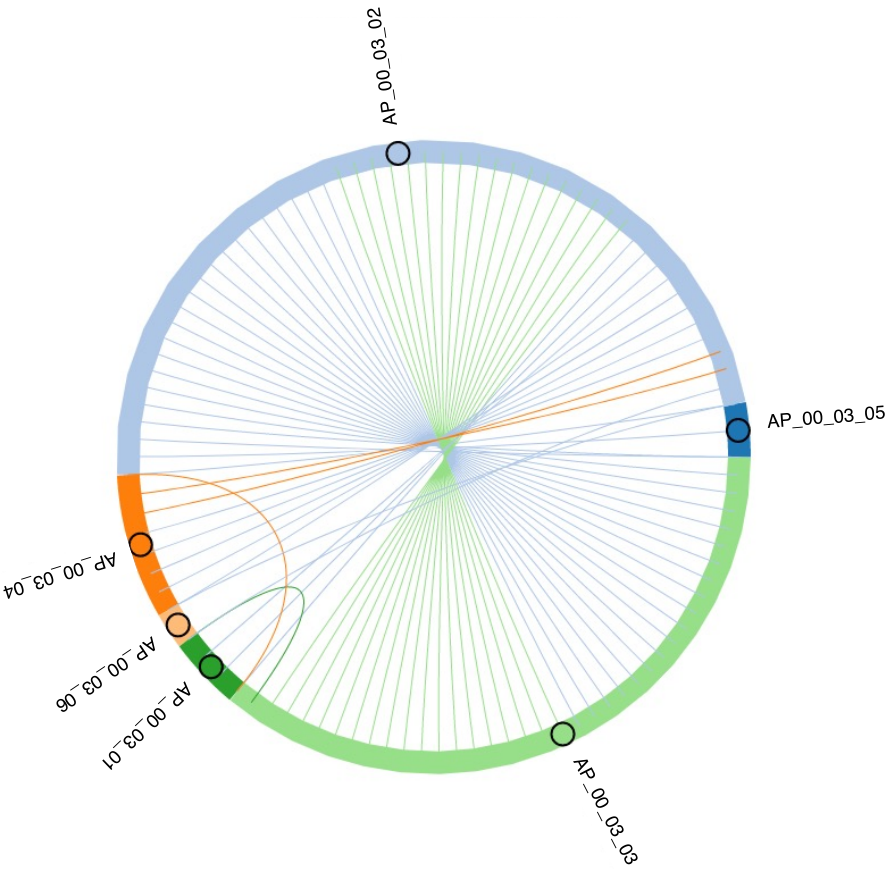}
        }
    \caption{Chord diagrams for three examples of the regions under study.}
    \label{fig:chord}
\end{figure}

Additionally, the chord diagrams of the transition matrices are generated to improve the interpretability of the models. Figure~\ref{fig:chord} presents the chord diagrams of the three examples presented in the previous paragraph, where the arcs represent the flow of the users between zones of a region, and the sizes of the arcs represent the frequency (probability) of the flows.

\subsection{Generation of the synthetic traces}

Once the user mobility model is obtained, it can be incorporated in a simulation/emulation tool. There are two alternatives, the incorporation of the data of the model (matrices and vectors) or the creation of a synthetic trace log that is read by the simulator. 

Our mobility model leaves the simulator designer the responsibility of defining the user arrival rate and the popularity of the user types. Thus, the simulator designer is able to change the \textit{workload} of the system in terms of the number of users. 

We implemented a Python script that randomly generates users and assigns them to a user type. For each new random user arrival, the user track is generated with the transition matrix and the time of the handoffs is determined with the time vector.

\subsection{Result analysis}

The analysis of the results consists of two-fold validations that test the correctness of the obtained model and its complexity. 

The first, in Section~\ref{sect_datamodelvalidation}, checks if the model generated with our method reflects accurately the movement of the users. For this case, the comparison with other previous works is not required, because we can compare the resulting model with the real raw data. In any case, we do not either compare our method with other research works because of our hierarchical definition, that is not studied in any previous works. Comparison with non-hierarchical models is not possible because the transition matrices have different shapes (different numbers of matrices and matrix sizes) and meaning.

Secondly, in Section~\ref{sect_hierarchicalvalidation}, the complexity and extrapolation of the resulting hierarchical model is evaluated and compared with a previous proposed non-hierarchical model (the Origin-Destination matrix)~\cite{BarbosaFilho2018HumanMM}.



\subsubsection{Mobility model analysis}
\label{sect_datamodelvalidation}

We have designed an experiment to analyze the quality of our model generation that consists of obtaining a second mobility model considering the synthetic traces generated with the first model. The greater the statistical similarity between both models, the best accuracy our proposal shows. We compute, and analyze, this statistical similarity with the mean square error (RMSE) over the transition matrices and the time vectors of each user. 

The complete process for the analysis of the case study is: (i) the gathering of real log traces; (ii) the generation of the mobility model of those traces; (iii) the creation of synthetic traces from the mobility model obtained; and (iv) the generation of the mobility model of these latter synthetic traces. Finally, the RMSE is calculated between the models obtained in (ii) and (iv).

The user types in (ii) and (iv) are mapped and the RMSE is calculated between the matrix/vector in each user type of (ii) and its corresponding matrix/vector in (iv). 
Table~\ref{tab:rmsedata} shows the average and maximum value of the RMSE for both the transition matrix and the time vector. Additionally, the error percentages are also included between brackets.

\begin{table}[h]
    \centering
    \footnotesize
    \begin{tabular}{l|cc|cc}
      \hline
     Region & \multicolumn{2}{c|}{Transition matrix} & \multicolumn{2}{c}{Time vector (s.)}\\
     & avg.  & max.  &  avg.  & max.  \\
     &  RMSE & RMSE &  RMSE &  RMSE \\
     \hline
    \cslzero & 0.025 (2.5\%) & 0.057 (5.7\%)  & 25.449 (4.5\%) & 62.207 (11.1\%) \\ 
    \hline
    \cslone & 0.077 (7.7\%) & 0.098 (9.8\%)  & 25.634 (16.8\%) & 31.651 (20.8\%) \\ 
    \hline
    \csltwo & 0.015 (1.5\%) & 0.038 (3.8\%) & 6.268 (5.6\%) & 17.826 (16.0\%) \\ 
    \hline

    \end{tabular}
    \caption{Maximum and mean value of the RMSE for the matrices and vectors of each region under study.}
    \label{tab:rmsedata}
\end{table}

The error levels are smaller in the case of the transition matrix, whose average values range between 7.7\% and 1.5\% and they are always smaller than 9.8\%. In the case of the time vector, the average error level ranges between 4.5\% and 16.8\%, and the maximum RMSE is 20.08\%. Note that the second region under study, \cslone, is the one with highest differences between both mobility models, and the case that increases the general error levels.

In general terms, we can conclude that the obtained model is more accurate in terms of the flow of the users between zones than in the time spent in a given zone.

\subsubsection{Hierarchical design analysis}
\label{sect_hierarchicalvalidation}

Our mobility characterization method is based on hierarchical modeling to minimize the impact of large scale scenarios. This hierarchical subdivision of the problem reduces the computational cost of subsequent analysis' tasks and even enables its distributed processing.  To answer the RQ1, which states  that our model reduces the complexity of the resulting model, and the execution time to obtain it, we have compared our hierarchical model with a non-hierarchical model, such as the one defined in  previous related works, the Origin-Destination matrix~\cite{BarbosaFilho2018HumanMM}. 

\begin{figure}[!h]
    \centering
       \includegraphics[width=0.45\textwidth, page=1]{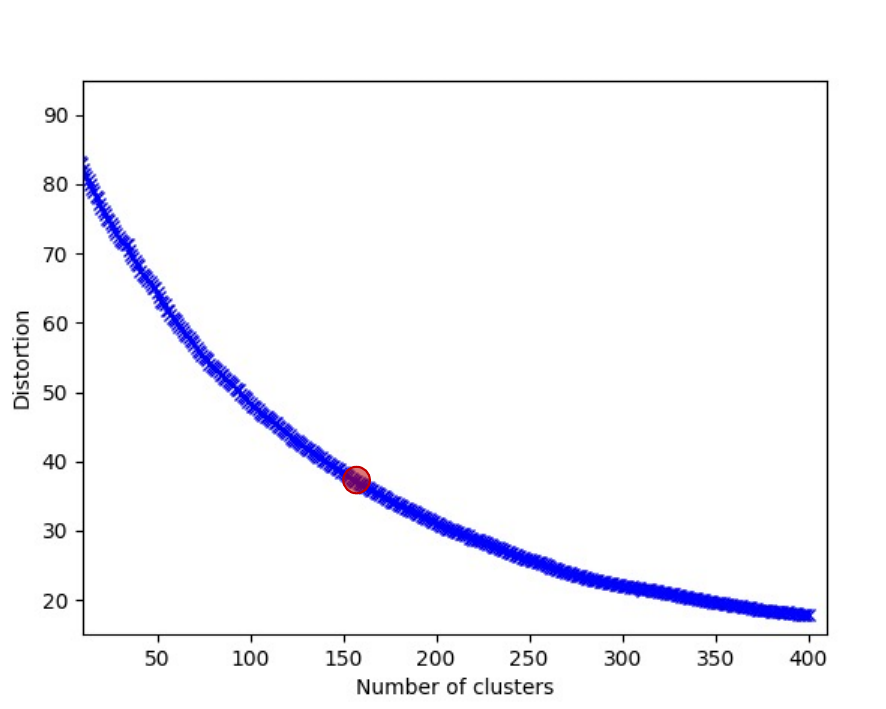}
    \caption{Elbow method to detect the optimal number of clusters considering the distortion metric and with a maximum number of 400 clusters.}
    \label{fig:elbow_all}
\end{figure}

Figure~\ref{fig:elbow_all} shows the number of optimal clusters for a non-hierarchical process in which each AP of our case study is mapped with a different zone. This is equivalent to one-level modeling with 425 zones. It is observed that the optimal number of clusters is much higher in the non-hierarchical modeling, resulting in 158 clusters, in opposition to the maximum number of 10 clusters across all the levels and regions in the hierarchical modeling. A smaller number of clusters reduces the complexity of the model and the execution times of the modeling process, and we have observed a very significant reduction from 158 to 10 clusters.


Moreover, the \textit{elbow} point of the non-hierarchical model (Figure~\ref{fig:elbow_all}) is less clear than the ones in the hierarchical model (Figure~\ref{fig:elbow}). The graph in the non-hierarchical case has a slope, or gradient, closer to -1.0. This means that an increase in the number of clusters could improve the distortion without a high increase of the cost/complexity of the clustering, for this non-hierarchical model.

As it has been commented in the previous paragraph, a higher number of clusters involves higher execution times. Table~\ref{tab:executiontimes} shows the execution times for the three examples of the case study and the non-hierarchical modeling. The total execution time is presented along with the desegregated time for the calculation of the optimal number of clusters, and the clustering algorithm. As it is observed, the execution time has an important improvement for the case of hierarchical modeling.  In general terms, the common execution time for each region of our case study is around 25 seconds, except for the case of the level 0, which increases up to 198 seconds because it is the only one with more than 10 zones. Considering that the hierarchy modeling of the campus defined 18 areas of level 1 and 74 areas of level 2, the execution time for all the regions under study is around 2500 seconds, five times shorter. Additionally, the execution could be parallelized very easily.



\begin{table}[h]
    \centering
    \begin{tabular}{l|rrr}
      \hline
     Region & \multicolumn{3}{c}{Execution time (s.)} \\
     & Total  & Elbow  & Clustering \\
     
     \hline
     Non-hierarchical & 14119.22& 13816.04& 30.64\\
    \cslzero & 197.95 & 157.86  & 1.97 \\
    \hline
    \cslone & 25.48 & 3.22 &  0.17 \\
    \hline
    \csltwo & 23.89 & 2.74 & 0.15\\
    \hline

    \end{tabular}
    \caption{Execution times for the modeling process.}
    \label{tab:executiontimes}
\end{table}

Finally, the goodness of the non-hierarchical modeling is measured with the RMSE (Table~\ref{tab:rmsenonhierarchical}). If we compare the RMSE with the ones of the hierarchical modeling (Table~\ref{tab:rmsedata}), it is observed that, although the average RMSE is better in the case of the transition matrices of the non-hierarchical modeling, the maximum value of the RMSE is similar to the worse case in the hierarchical modeling. On the contrary, the error level for the time vector is better for the case of the non-hierarchical modeling.

\begin{table}[h]
    \centering
    \footnotesize
    \begin{tabular}{l|cc|cc}
      \hline
      & \multicolumn{2}{c|}{Transition matrix} & \multicolumn{2}{c}{Time vector (s.)}\\
     & avg.  & max.   & avg.  & max.   \\
     &  RMSE & RMSE &  RMSE &  RMSE \\
     \hline
    Non-hierarchical & 0.019 (1.9\%) & 0.088 (8.8\%)  & 4.981 (0.6\%) & 25.475 (3.1\%) \\ 
    \hline
    \end{tabular}
    \caption{Maximum and average value of the RMSE for the non-hierarchical modeling.}
    \label{tab:rmsenonhierarchical}
\end{table}

To sum up, the hierarchical modeling shows similar error levels in the transition matrices, but with a worse behavior in the time vector. The great benefit of a hierarchical method is in the complexity both in the number of clusters and the execution time of the modeling.


The second research gap that our proposal deals with is to allow the extrapolation or adaptability of the resulting models between scenarios with different geospatial features. To answer to this second research question (RQ2), we have extrapolated the mobility model obtained in this case study. 
Remember that the analysis of the results of our case study are focused on one building (\textit{bldgAT}) in a university campus with 18 buildings. We have extrapolated this real scenario to a fictitious one with only 3 buildings with similar features to \textit{bldgAT}. To do that, we replicate the model of \textit{bldgAT}, and we adapt each of those three replicated transition matrices to create buildings with different number of winds and floors. 

The new modified model is used to generate synthetic traces. An additional modeling process is applied to compare the results with the former model of \textit{bldgAT}. The results show that the average RMSE are 2.85\% and 4.31\% respectively for the transition matrix and the time vector of \textit{bldgAT}.

\section{Conclusions}
\label{sect_conclusions}


We have proposed a method for user mobility characterization based on a hierarchical definition of regions to (RQ1) reduce the complexity of creating a mobility model based on transition matrices, and to (RQ2) increase the adaptability of the obtained models between scenarios with different geospatial features.  The applicability of the method has been tested in a case study located in the campus of the \uiblong.

The case study was analyzed in terms of the RMSE between the model obtained from the real data traces and the model obtained from the synthetic traces created with the former model. Additionally, a non-hierarchical study of the campus was also performed.

By comparing the results between the hierarchical and the non-hierarchical study, we observed that the complexity of the modeling process is much lower in the first one,  which answered the RQ1. Although the non-hierarchical model showed smaller values for the RMSE, both studies showed acceptable error levels.  Additionally, RQ2 was answered by extrapolating the obtained hierarchical model to a scenario with different geospatial features. The resulting model was used to create synthetic mobility which showed low RMSE values.

We propose three future research lines that emerge from the proposal of our method: the integration of geospatial hierarchical mobility models in a fog simulator, such as iFogSim~\cite{10.1002/spe.2509} or YAFS~\cite{8758823}; the extension of other mobility models, instead of the ones that use a transition matrix, with a geospatial hierarchical design; and, finally, the adaptation of the proposed method to other problem scenarios, such as the prediction of the user behavior.

In terms of the case study, an interesting future work is to apply our method to a case study where the user mobility data is gathered with a different wireless technology, for example a 5G network instead of a Wi-Fi. This would prove that our method is not influenced by the underlying network technology that is used to gather user mobility data. 
The main difficulty to carry it on in the future is to get access to these types of data, which are only available in the infrastructures of telecommunication companies.





\section*{Data statement}
The data generated in this study and the source code of the scripts are public available in the repository \url{\github}.

\section*{Acknowledgements}
Funding: Project TIN2017-88547-P supported by the Spanish Goverment (MCIN/ AEI /10.13039/501100011033/) and the European Commission (FEDER Una manera de hacer Europa).




\bibliographystyle{elsarticle-num} 
\bibliography{mybib}





\end{document}